\documentclass[aps,onecolumn,11pt,floatfix,altaffilletter,tightenlines,showpacs,showkeys,notitlepage,nofootinbib]{revtex4-1}
\usepackage[dvipdf,dvips,dvipdfmx]{graphicx}
\usepackage[colorlinks=true,citecolor=blue,linkcolor=blue]{hyperref}
\usepackage{amsmath}
\usepackage{amssymb}
\usepackage{epsfig}
\usepackage{epstopdf}
\usepackage{url}
\usepackage{color}
 \usepackage[utf8]{inputenc} 

\newcommand{\Slash}[1]{{\ooalign{\hfil/\hfil\crcr$#1$}}} 

\newcommand{\nn}{\nonumber}
\newcommand{\be}{\begin{eqnarray}}
\newcommand{\ee}{\end{eqnarray}}
\newcommand{\al}[1]{\begin{align}#1\end{align}}
\catcode`\@=11

\def\lsim{\mathrel{\mathpalette\@versim<}}
\def\gsim{\mathrel{\mathpalette\@versim>}}
\def\@versim#1#2{\vcenter{\offinterlineskip
\ialign{$\m@th#1\hfil##\hfil$\crcr#2\crcr\sim\crcr } }}
\catcode`\@=12

\begin{document}

\title{Gravitational waves \\ from chiral phase transition\\
in a conformally extended standard model}

\author{Mayumi  Aoki$^{1\,}$} \email{mayumi@hep.s.kanazawa-u.ac.jp}
\author{Jisuke Kubo$^{2,3\,}$} \email{jikubo4@gmail.com}

\affiliation{%
        $^1$Institute for Theoretical Physics, Kanazawa University, 
        Kanazawa 920-1192, Japan\\
 $^2$Max-Planck-Institut f\"ur Physik (Werner-Heisenberg-Institut),
F\"ohringer Ring 6, D-80805 M\"unchen, Germany   \\
	$^3$Department of Physics, University of Toyama, 3190 Gofuku, Toyama 930-8555, Japan%
}

\preprint{KANAZAWA-19-07}
\preprint{MPP-2019-200}

%
\begin{abstract} 
The gravitational wave (GW)  background produced 
at the cosmological chiral phase transition  in 
a conformal extension of the standard model is studied.
To obtain the bounce solution of coupled field equations
we implement an iterative method.  We find that
the corresponding $O(3)$ symmetric 
Euclidean action 
 $S_3$ divided by the temperature $T$ has a simple behavior near the
critical temperature $T_C$: $S_3/T \propto
 (1-T/T_C)^{-\gamma}$, which is subsequently used
to determine the transition's inverse duration 
$\beta$ normalized to the Hubble parameter $H$.
It turns out that $\beta/H \gsim 10^3$, implying
that  the sound wave period $\tau_\text{sw}$ as  an active 
GW source, too,  can be much shorter than the Hubble time. 
We therefore compute
$\tau_\text{sw} H$ and  use it as  the reduction 
factor for the sound wave contribution.
The signal-to-noise ratio (SNR) for Deci-Hertz Interferometer Gravitational Wave Observatory (DECIGO)  and Big Bang Observer (BBO) is evaluated, with the result:
 SNR$^\text{DECIGO} \lsim 1.2$ and 
SNR$^\text{BBO} \lsim 12.0$ for five years observation,
from which we conclude  that
the GW signal predicted by  the model in the optimistic case could be detected at BBO.

\end{abstract}
\maketitle

%
\section{Introduction}

One of the central questions in particle physics today  is:
How to go beyond the standard model (SM), see, e.g.,
\cite{Lindner2019}.
Indeed many theoretical suggestions have been made since ever
\cite{Ross:2017oiz}.
The fact that the Higgs mass term is the only dimensionful
parameter in the SM
and the theory is perturbative -- no Landau pole 
below the Planck scale \cite{Holthausen:2011aa,Bezrukov:2012sa,Degrassi:2012ry,
Buttazzo:2013uya} --
may be regarded as a hint of how to go beyond the SM \cite{Lindner2019}.
Even before the SM was proposed,
John Wheeler \cite{Wilczek:1999be}
 wished to remove 
all the dimensionful parameters from the fundamental equations. 
If we start with a 
theory, which at the classical level contains  no dimensionful parameter
such as mass parameter at all, an energy scale  has to be generated 
by quantum effects.
A quantum generation  of the Higgs mass term
from ``nothing'' 
would be along the line of John Wheeler's thought.
There are two known mechanisms of ``scalegenesis'': One is the Coleman--Weinberg mechanism \cite{Coleman:1973jx}
that is based on improved perturbation theory and works thanks
to scale anomaly
\cite{Callan:1970yg,Symanzik:1970rt}.
The other one is the dynamical scale 
symmetry breaking by  strong dynamics  in nonabelian 
gauge theories, e.g., Quantum Chromodynamics (QCD).
We recall  that about 99 \% of the energy portion
of the ordinary matter in the Universe -- baryon -- is generated by the 
nonperturbative effect in QCD \cite{Kronfeld:2012ym}, 
dynamical chiral symmetry 
breaking \cite{Nambu:1960xd,Nambu:1961tp,Nambu:1961fr}.
Several realistic models  using the strong dynamics have been suggested
in \cite{Hur:2011sv,Heikinheimo:2013fta,Holthausen:2013ota,Kubo:2014ova,Kubo:2015cna,Hatanaka:2016rek}:
It has been found that not only the Higgs mass term,
but also
the dark matter mass \cite{Hur:2011sv,Heikinheimo:2013fta,Holthausen:2013ota,Kubo:2015cna,Hatanaka:2016rek,Kubo:2014ida,Ametani:2015jla}, contributing to 
27 \% of the total energy of the Universe \cite{Aghanim:2018eyx},
as well as the Planck mass \cite{Kubo:2018kho}
 can be  generated by dynamical scale symmetry breaking.

At finite temperature the  real QCD does not undergo  
a phase transition (PT),  
rather a continuous change of crossover
type \cite{Bhattacharya:2014ara}. 
However, for  sufficiently small
current quark masses, the system  can undergo a first-order PT
\cite{Pisarski:1983ms, DeTar:2009ef, Meyer:2015wax,Jin:2017jjp}, and such a situation can be realized
in  hidden sector models
\cite{Holthausen:2013ota,Kubo:2015joa,Aoki:2017aws} 
(see also \cite{Schwaller:2015tja} and references therein),
 in which dynamical  breaking  of scale symmetry takes place at energies
 higher than the SM scale.
 If the coupling of the hidden sector to the SM is very small,
a  chief signal from the hidden sector is
 the gravitational wave (GW) background produced at a first-order 
 PT in a certain epoch of the Universe \cite{Witten:1984rs},
  see e.g. refs. \cite{Maggiore:1999vm,Binetruy:2012ze} for reviews.
  \footnote{
	The  crossover transition in the real QCD
	can influence the  spectrum of the inflationary GW
	\cite{Seto:2003kc,Kuroyanagi:2008ye,Schettler:2010dp,Schettler:2010wi,Saikawa:2018rcs,Hajkarim:2019csy}.
	The  frequency band of the damped GWs is 
	what has been predicted by  Witten \cite{Witten:1984rs}.
}
This has been even more the case
 since the GWs have been  detected on the earth 
\cite{Abbott:2016blz,TheLIGOScientific:2017qsa,GBM:2017lvd}.

In this paper we consider the model 
\cite{Kubo:2014ida,Ametani:2015jla}, in which
a robust energy scale, created by 
the chiral symmetry breaking 
in a strongly interacting QCD-like hidden sector,
transmits via a SM singlet real scalar mediator  $S$ to the SM sector
and generates the Higgs mass term to trigger
 electroweak (EW) symmetry breaking.
We are particularly interested in the GW background produced
at the cosmological chiral PT of the model.
\footnote{The GWs produced during a cosmological first-oder PT
in  classically scale invariant models have been recently 
studied in refs. \cite{Konstandin:2011dr,Hashino:2015nxa,Tsumura:2017knk,Marzola:2017jzl,Jinno:2016knw,Prokopec:2018tnq,Hashino:2018wee,Brdar:2018num,Miura:2018dsy,Marzo:2018nov,Croon:2019iuh,Dai:2019ksi,Mohamadnejad:2019vzg}.}
The present work is an extension  of ref. \cite{Aoki:2017aws},
where a few benchmark points in the parameter space have
been chosen to study
the GW background spectrum. We have decided to extend the analysis of 
ref. \cite{Aoki:2017aws} from the following reasons:\\
a) 
The GW energy density depends strongly on the ratio
of the duration time $\tau_\text{PT}=1/\beta$ of the first-order PT 
to the Hubble time $1/H$, i.e., $(\beta/H)^{-1}$
\cite{Witten:1984rs,Kosowsky:1991ua,Kosowsky:1992rz,Kosowsky:1992vn}. 
Using effective field theories it  has been shown \cite{Helmboldt:2019pan}
that, in contrast to  the commonly assumed value
of $\beta/H\sim O(10^2)$ \cite{Hogan:1984hx,Witten:1984rs}
(see also \cite{Maggiore:1999vm}), 
it is of order $10^4$  in QCD like theories if the coupling to 
the SM is neglected, i.e., in the absence of the mediator $S$. This means a large suppression of the GW energy density.
Here we will systematically look for a parameter space with
smaller $\beta/H$, which leads to larger GW energy densities.\\
b) It turns out that the influence of the mediator $S$ is an important 
factor to decrease $\beta/H$; 
the quartic self-coupling of $S$, $\lambda_S$, 
should be of order $10^{-3}$, which is much smaller than the Higgs
self-coupling $\lambda_H\sim O(10^{-1})$.
Consequently, the mass of $S$ denoted by $m_S$ 
can become comparable with -- or even smaller than -- the Higgs mass $m_h$, and
consequently 
the mixing of the Higgs $h$ and $S$ is no longer negligible,
i.e., subject to the LHC constraint (see e.g. refs. 
\cite{Falkowski:2015iwa,Robens:2016xkb}).
We will here take into account this LHC constraint.\\
c)
To compute $\beta/H$ one has to solve  classical equations of motion
and  obtain the so-called bounce solution that describes a bubble
appearing during   a first-order PT \cite{Linde:1981zj}.
In the model in question there are two fields that are involved in the
problem, $\sigma$ for the chiral condensate and $S$, so that 
we have to deal with a system of  coupled  differential equations.
In ref. \cite{Aoki:2017aws} 
we have employed a (modified) path-deformation method
\cite{Wainwright:2011kj} to solve them. 
However, it has turned out that this method suffers from a large uncertainty and does not yield trustful results. 
 Here we will employ another iterative method
to realize a faster convergence of the iterative process.\\
d) The sound wave contribution to the GW spectrum
will be the most dominant contribution in the model we will consider.
A large $\beta/H$ means a short duration of the first-order 
cosmological PT
and hence a short sound wave period $\tau_\text{sw}$
compared with $1/H$. However, 
the formula for the sound wave contribution to
the GW spectrum has been derived 
from the numerical simulations for which a long-lasting 
source of the GW, i.e., $\tau_\text{sw}H > 1$, is assumed 
\cite{Hindmarsh:2017gnf}.
If  $\tau_\text{sw}H <  1$,
 the sound wave is an active GW source only for a period shorter than the Hubble time. The above-mentioned formula therefore overestimates the sound wave contribution.
Following refs. \cite{Ellis:2018mja,Ellis:2019oqb} along with ref. 
\cite{Espinosa:2010hh}, 
we calculate  $\tau_\text{sw}H $
and use it  as the reduction factor for the sound wave contribution.\\
 e)
The signal-to-noise ratio (SNR) is an important measure 
to evaluate the detectability of the GW background of the model
\cite{Thrane:2013oya}. We will calculate the SNR
for Deci-Hertz Interferometer 
Gravitational Wave Observatory (DECIGO) \cite{Seto:2001qf,Kawamura:2006up,Kawamura:2011zz}
and Big Bang Observer (BBO) \cite{BBO,Cornish:2005qw,Corbin:2005ny}.

In Section II we  outline the basic feature of the model;
dynamical generation of the Higgs mass term,
mass spectrum,  
the LHC constraint of the Higgs-$S$ mixing,
and dark matter (DM).
Since the hidden sector of the model is strongly interacting, we  use
an effective theory for the dynamical chiral symmetry breaking --
the  Nambu-Jona-Lasinio (NJL) model 
\cite{Nambu:1960xd,Nambu:1961tp,Nambu:1961fr} -- as in refs. 
\cite{Holthausen:2013ota,Kubo:2014ida,Ametani:2015jla,Aoki:2017aws}, where
our approximation method, 
the self-consistent mean-field approximation 
(SCMF) of refs. \cite{Kunihiro:1983ej,Hatsuda:1994pi}, 
is also briefly elucidated in this section.

After a short review on the chiral PT in the hidden sector of the model
we present, in Section III, our iterative 
method to obtain the bounce solution.
We narrow the  parameter space with smaller $\beta/H$.
Two benchmark points are chosen  for an orientation
of the parameter space that we consider.
In Section IV we discuss the GW  spectrum.
The  above-mentioned reduction factor $\tau_\text{sw}H$ for the sound wave contribution
is computed in this section.
We then calculate the SNR to evaluate the detectability of the GW signal 
at DECIGO and BBO.  We also compare
the GW spectrum for  two chosen benchmark points
with the power-law integrated sensitivity 
\cite{Thrane:2013oya} of  DECIGO and BBO.
Section V is devoted to summary and conclusion.

\section{The Model}\label{Hidden QCD Physics}
We consider a classically scale invariant extension of the SM studied in 
refs. \cite{Kubo:2014ida,Ametani:2015jla}. 
The model consists of a hidden $SU(n_c)_H$ gauge sector coupled
to the SM sector via a real singlet scalar $S$.
The hidden sector Lagrangian ${\cal L}_{\rm H}$ of the total Lagrangian 
${\cal L}_T ={\cal L}_{\rm H}+{\cal L}_{\mathrm{SM}+S}$ of the model 
is given as 
\be
{\cal L}_{\rm H}
&=-\frac{1}{2}\mbox{Tr}~F^2+
\mbox{Tr}~\bar{\psi}(i\gamma^\mu \partial_\mu +
g_H \gamma^\mu G_\mu +g'  Q \gamma^\mu B_\mu-
{\bf y} S)\psi \,,
\label{eq:LH}
\ee
where $G_\mu$ is the gauge field for the hidden QCD,
$B_\mu$ is the $U(1)_Y $ gauge field, 
\begin{align}
B_\mu &=\cos \theta_W A_\mu-\sin\theta_W Z_\mu~,~~
g'=e/\cos \theta_W \,,
\end{align}
and the hidden vector-like fermions $\psi_i~(i=1,\dots,n_f)$ 
belong to the fundamental representation
of  $SU(n_c)_H$.
The  ${\bf y}$ is an $n_f\times n_f$ Yukawa coupling matrix which can be taken as a diagonal matrix without loss of generality, i.e. 
${\bf y}=\mbox{diag.}(y_1,\dots,y_{n_f})$.
Here the diagonal entries  $y_i$ are assumed to be positive.
The ${\cal L}_{\mathrm{SM}+S}$ part contains the 
SM gauge and Yukawa  interactions
along with the scalar potential
\begin{align}
V_{\mathrm{SM}+S}
&=
\lambda_H ( H^\dag H)^2+
\frac{1}{4}\lambda_S S^4
-\frac{1}{2}\lambda_{HS}S^2(H^\dag H)\,,
\label{eq:VSM}
\end{align}
where the portal coupling $\lambda_{HS}$ is assumed to be positive,
and  $H^T=( H^+,~(h+iG^0)\sqrt{2}  )$ is the SM Higgs doublet field
with $H^+$ and $G^0$ as the would-be Nambu-Goldstone (NG) fields.
The (tree-level) stability condition for the scalar potential is given by
\begin{align}
\label{stability}
    \lambda_{H}>0,~~\lambda_{S}>0, ~~2\sqrt{\lambda_{H}\lambda_{S}}-\lambda_{HS}&>0 \,.
\end{align}

Following refs. \cite{Holthausen:2013ota,Kubo:2014ida,Ametani:2015jla} we consider $n_f=n_c=3$.
In this case, the hidden chiral symmetry 
$\mathrm{SU}(3)_{L}\times\mathrm{SU}(3)_{R}$ is dynamically broken to its diagonal subgroup $SU(3)_V$ by the nonzero chiral condensate
$\left<\bar{\psi}\psi\right>$,
which implies the existence of 8 NG bosons.
At the same time of the dynamical chiral symmetry breaking, the singlet scalar field $S$ 
acquires a nonzero vacuum expectation value (VEV)  due to the Yukawa
interaction $-yS\bar{\psi}\psi$ in  ${\cal L}_{\rm H}$,  generating an
explicit-chiral-symmetry-breaking mass term.
Consequently, the NG bosons acquire their masses and can become  DM candidates due to the remnant unbroken flavor group $SU(3)_V$
(or its subgroup, depending on the choice of $y_i$) that can stabilise them.
Finally, with the nonzero $v_S=\left<S\right>$, 
the EW symmetry breaking is triggered by the Higgs mass term  $+\frac{1}{2}\lambda_{HS}v_S^2H^{\dag}H$.

\subsection{Nambu--Jona-Lasinio description}
In order to analyze the strongly interacting hidden sector,
we replace the Lagrangian ${\cal L}_{\rm H}$  (\ref{eq:LH}) by
  the  NJL Lagrangian 
 that serves as an effective Lagrangian for the dynamical chiral symmetry breaking
 \cite{Nambu:1960xd,Nambu:1961tp,Nambu:1961fr}: 
 \begin{align}
{\cal L}_{\rm NJL}&=\mbox{Tr}~\bar{\psi}(i\gamma^\mu\partial_\mu 
+g' Q \gamma^\mu B_\mu-{\bf y} S)\psi+2G~\mbox{Tr} ~\Phi^\dag \Phi
+G_D~(\det \Phi+h.c.)\,,
\label{eq:NJL10}
\end{align}
where
\begin{align}
\Phi_{ij}&= \bar{\psi}_i(1-\gamma_5)\psi_j=
\frac{1}{2}\sum_{a=0}^{8}
\lambda_{ji}^a\, [\,\bar{\psi}\lambda^a(1-\gamma_5)\psi\,]\,,
\end{align}
and $\lambda^a (a=1,\dots, 8)$ are the Gell-Mann matrices with
$\lambda^0=\sqrt{2/3}~{\bf 1}$.
The dimensionful parameters $G$ and $G_D$ have canonical dimensions of $-$2 and $-$5, respectively.
In order to deal with the nonrenormalizable Lagrangian (\ref{eq:NJL10}) we work in 
the SCMF approximation
of refs. \cite{Kunihiro:1983ej,Hatsuda:1994pi}.
The mean fields $\sigma_i~(i=1,2,3)$ 
and $\phi_a~(a=0,\dots,8)$ are defined in the ``Bardeen-Cooper-Schrieffer" vacuum as
\begin{align}
\label{varphi}
\sigma_i =- 4 G\left<\bar{\psi}_i \psi_i  \right> \,, ~~~
\phi_a =-2 i G\left<\bar{\psi}_i \gamma_5 \lambda^a\psi_i  \right>\,,
\end{align}
where the CP-even mean fields corresponding 
to the non-diagonal elements of $\langle \bar{\psi}_i\psi_j \rangle$ are suppressed, because they do not play any role
for our purpose.
Splitting the NJL Lagrangian $\mathcal{L}_{\text{NJL}}$ into two parts as 
 $\mathcal{L}_{\text{NJL}} =\mathcal{L}_{\text{MFA}}+\mathcal{L}_{I}$ 
 where $\mathcal{L}_{I}$ is normal ordered (i.e., $\langle 0|\mathcal{L}_{I}|0\rangle =0$),
 we find the Lagrangian in the SCMF approximation $\mathcal{L}_{\text{MFA}}$
 in the $SU(3)_V$  limit  as
 \footnote{The mean-field Lagrangian $\mathcal{L}_{\text{MFA}}$
 in the case of broken $SU(3)_V$ can be found in ref. \cite{Ametani:2015jla}.}
\begin{align}
\nonumber
   \mathcal{L}_{\text{MFA}} = &   
   \mathrm{Tr} ~\bar{\psi}(i\Slash{\partial}-M+g' Q \gamma^\mu B_\mu)\psi -i\mathrm{Tr} ~\bar{\psi}\gamma_5 \phi \psi -\frac{1}{8G}\left( 3\sigma^2+2\sum^8 _{a=1} \phi_a \phi_a \right)  \\
 \label{Hidden SCMFA}
    & +\frac{G_D}{8G^2}\left(  -\mathrm{Tr} ~\bar{\psi} \phi^2 \psi + \sum^8 _{a=1} \phi_a \phi_a \mathrm{Tr} ~\bar{\psi}\psi + i\sigma \mathrm{Tr}~ \bar{\psi}\gamma_5 \phi \psi +\frac{\sigma^3}{2G}+\frac{\sigma}{2G}\sum^8 _{a=1} (\phi_a)^2   \right) 
\end{align}
with $\phi=\sum_{a=1}^8~\phi_a \lambda^a$ and $\sigma=\sigma_1=
\sigma_2=\sigma_3$.
Here
$\phi_0$ has been suppressed and 
the constituent fermion mass $M$ is given by 
\begin{align}
\label{M}
M(S,\sigma)= \sigma+yS-\frac{G_D}{8G^2}\sigma^2\,,
\mbox{where}~y=y_1=y_2=y_3\,.
\end{align}

The one-loop effective potential obtained from $\mathcal{L}_{\text{MFA}}$
(\ref{Hidden SCMFA}) 
can be obtained by integrating out the hidden fermions:
\begin{align}
V_{\rm NJL}(S,\sigma)
& = \frac{3}{8G}\sigma^2-
\frac{G_D}{16G^3}\sigma^3
-3n_c I_0(M,\Lambda_H)\,.
\label{eq:Vnjl}
\end{align}
Here the  function $I_0$ is given by
\begin{align}
  I_0(M, \Lambda)= \frac{1}{16\pi^2}\left[ \Lambda^4 \ln \left( 1+\frac{M^2}{\Lambda^2 }\right)-M^4 \ln \left( 1+\frac{\Lambda^2 }{M ^2}\right) + \Lambda^2 M ^2\right] 
\end{align}
with a four-dimensional momentum cutoff $\Lambda$,
where we denote the cutoff in the hidden sector
by $\Lambda_H$.
For a certain interval of  the dimensionless 
parameters $G^{1/2}\Lambda_H$ and $(-G_D)^{1/5}\Lambda_H$
we have $\left<\sigma\right>\ne 0$ and $\left<S \right>\ne 0$
\cite{Holthausen:2013ota,Kubo:2014ida,Ametani:2015jla}. 
It is then meant that the dynamics of the hidden sector 
creates a nonvanishing chiral condensate 
$\left<0|\bar\psi_i \psi_i|0\right>\ne 0$.
One can see that 
the potential $V_{\rm NJL}(S,\sigma)$ is asymmetric in $\sigma$ owing to the last term in the NJL Lagrangian (\ref{eq:NJL10}) and also from the constituent mass $M$ (\ref{M}), which is the reason that the chiral PT at finite temperature
can become of first order.
It is noted that the mean fields $\sigma$ and $\phi_a$ are non-propagating classical fields at the tree level.
Therefore, their kinetic terms are generated by integrating out the hidden fermions at the one-loop level, which will be seen in Section \ref{Mass spectrum} where 
two point functions are calculated.

The NJL parameters for the hidden QCD sector are obtained by scaling-up the values of
$G, G_D$ and the cutoff $\Lambda$ from QCD hadron physics. 
Following refs.
\cite{Holthausen:2013ota,Kubo:2014ida,Ametani:2015jla} we assume that the dimensionless combinations
\begin{align}
 G^{1/2}\Lambda_H=1.82\,,~~~~(-G_D)^{1/5}\Lambda_H=2.29\,,
 \label{NJL para}
\end{align}
which are satisfied for the real-world hadrons, remain unchanged for a higher scale of $\Lambda_H$. Therefore, the free parameters of the (effective) model
are: $\lambda_H, \lambda_S, \lambda_{HS}$ and $\Lambda_H$.
Once these parameters are fixed,
the VEVs of $\sigma\,, S$ and $h$
can be obtained through the minimization of the scalar potential 
$V_{\mathrm{SM}+S}+V_\text{NJL}$
where we choose these parameters so as to satisfy 
$m_h=125$ GeV and $\left<h\right>=246~\mathrm{GeV}$.

\subsection{Mass spectrum}\label{Mass spectrum}
Once the VEVs of $\sigma$, $S$ and $h$ are obtained, 
the scalar mass spectrum can be calculated from 
the corresponding two point functions at one-loop
in which  the hidden fermions are circulating.
The CP even scalars $h, S$ and $\sigma$ mix with each other.
The two point functions at the one-loop level $\Gamma_{AB} (A,B=h,S,\sigma)$
in the $SU(3)_V$ flavor symmetry limit are given by
\begin{align}
\Gamma_{hh}(p^2)&=p^2-3\lambda_{H}\left<h\right>^2+\frac{1}{2}\lambda_{HS}\left<S\right>^2\,, ~~
\Gamma_{hS}=\lambda_{HS}\left<h\right>\left<S\right>\,,  ~~
\Gamma_{h\sigma}=0\,,  \nonumber \\
\Gamma_{SS}(p^2)&=p^2-3\lambda_{S}\left<S\right>^2+\frac{1}{2}\lambda_{HS}\left<h\right>^2-y^23n_cI_{\varphi^2}(p^2,M,\Lambda_{\mathrm{H}})\,, \label{GammaSS} \\
\Gamma_{S\sigma}(p^2)&=-y\left(1-\frac{G_D\left<\sigma\right>}{4G^2}\right)3n_cI_{\varphi^2}(p^2,M,\Lambda_{\mathrm{H}})\,, 
\label{GammaSs}\\
\Gamma_{\sigma\sigma}(p^2)&=-\frac{3}{4G}+\frac{3G_D\left<\sigma\right>}{8G^3}-\left(1-\frac{G_D\left<\sigma\right>}{4G^2}\right)^23n_cI_{\varphi^2}(p^2,M,\Lambda_{\mathrm{H}}) 
+\frac{G_D}{G^2}3n_cI_V(M,\Lambda_{\mathrm{H}})\,.\nn
\end{align}
Here the loop functions are defined as
\begin{align}
 \label{propagator for sigma}
  I_{\varphi^2}(p^2,M,\Lambda) &= \int_{\Lambda} \frac{d^4 k}{i(2\pi)^4}\frac{\mathrm{Tr}(\Slash{k}+\Slash{p}+M)(\Slash{k}+M)}{((k+p)^2-M^2)(k^2-M^2)}\,, \\
  I_{V}(M,\Lambda) & = \int_{\Lambda}  \frac{d^4 k}{i(2\pi)^4}\frac{M}{(k^2-M^2)}=-\frac{1}{16\pi^2}M\left[ \Lambda^2-M^2\ln \left( 1+\frac{\Lambda^2}{M^2}\right) \right]\,.
\end{align}
The flavor eigenstates $(h, S, \sigma)$ and the mass eigenstates $h_i ~(i=1, 2, 3)$ are related by 
\al{
\left(\begin{array}{c}h\\S\\\sigma
\end{array}\right) &=
\left(\begin{array}{ccc}
\xi_h^{(1)} & \xi_h^{(2)} & \xi_h^{(3)} \\ 
\xi_S^{(1)} & \xi_S^{(2)} & \xi_S^{(3)} \\ 
\xi_\sigma^{(1)} & \xi_\sigma^{(2)} & \xi_\sigma^{(3)} 
\end{array}  \right)~\left(\begin{array}{c}h_1\\h_2 \\h_3
\end{array}\right)\,.
\label{mixingM}}
The squared masses 
$m^2_{h_i}$ are determined by the zeros of the two point functions 
at the one-loop level,
i.e. $\Gamma_{AB}(m^2_{h_i})\xi_{B}^{(i)}=0$. 

In this model the DM candidates are the NG bosons in the hidden sector which are CP-odd scalars $\phi_a$ 
in eq. (\ref{varphi}), i.e. the dark mesons.
The two point function at the one-loop level for the DM candidate is 
(in the $SU(3)_V$ flavor symmetry limit)
\begin{align}
\label{GammaDM}
\Gamma_{\mathrm{DM}}(p^2)&=-\frac{1}{2G}+\frac{G_D\left<\sigma\right>}{8G^3}+\left(1-\frac{G_D\left<\sigma\right>} {8G^2}\right)^22n_cI_{\phi^2}(p^2,M,\Lambda_{\mathrm{H}}) +\frac{G_D}{G^2}n_cI_V(M,\Lambda_{\mathrm{H}})\,,
\end{align}
where the loop function $ I_{\phi^2}(p^2,M,\Lambda)$ is given by
\begin{align}
  I_{\phi^2}(p^2,M,\Lambda) &= \int_{\Lambda}  \frac{d^4 k}{i(2\pi)^4}\frac{\mathrm{Tr}(\Slash{k}-\Slash{p}+M)\gamma_5(\Slash{k}+M)\gamma_5}{((k-p)^2-M^2)(k^2-M^2)}\,.
\end{align}
The mass of the DM is obtained from 
$\Gamma_{\mathrm{DM}}(m_{\mathrm{DM}}^2)=0$.


\subsection{LHC constraint on $\lambda_{HS}$}
The size of the portal coupling $\lambda_{HS}$ controls
the $h-S$ mixing.
Since in the parameter space we will consider
the Yukawa coupling $y$ is small, i.e. of order $10^{-3}$ 
(see eq. (\ref{parameter})),  the mixing
$\Gamma_{S\sigma}$ (\ref{GammaSs}) is also small, so that
we will neglect it  in the following discussions.
(We will also neglect the last term of 
$\Gamma_{SS}$ (\ref{GammaSS}), because it is proportional 
to $y^2$.)
Therefore, the  $h-S$ mixing  can be written as
\al{
\left(\begin{array}{c}h_1\\h_2
\end{array}\right) &=
\left(\begin{array}{cc}\cos\theta & \sin\theta\\ -\sin\theta &\cos\theta
\end{array}  \right)~\left(\begin{array}{c}h\\S
\end{array}\right) }
with  $(\cos\theta\,,\,-\sin\theta\,,\,
0)\simeq (\xi_h^{(1)}\,,\,\xi_h^{(2)}\,,\,
\xi_h^{(3)})$ which is defined in eq. (\ref{mixingM}).
Here we identify $h_1$ with the SM Higgs having mass $m_h \simeq  0.125$ TeV,
i.e., $m_h=m_{h_1}$  and  $m_S=m_{h_2}$.
 \begin{figure}[t]
\begin{center}
\includegraphics[width=3.2in]{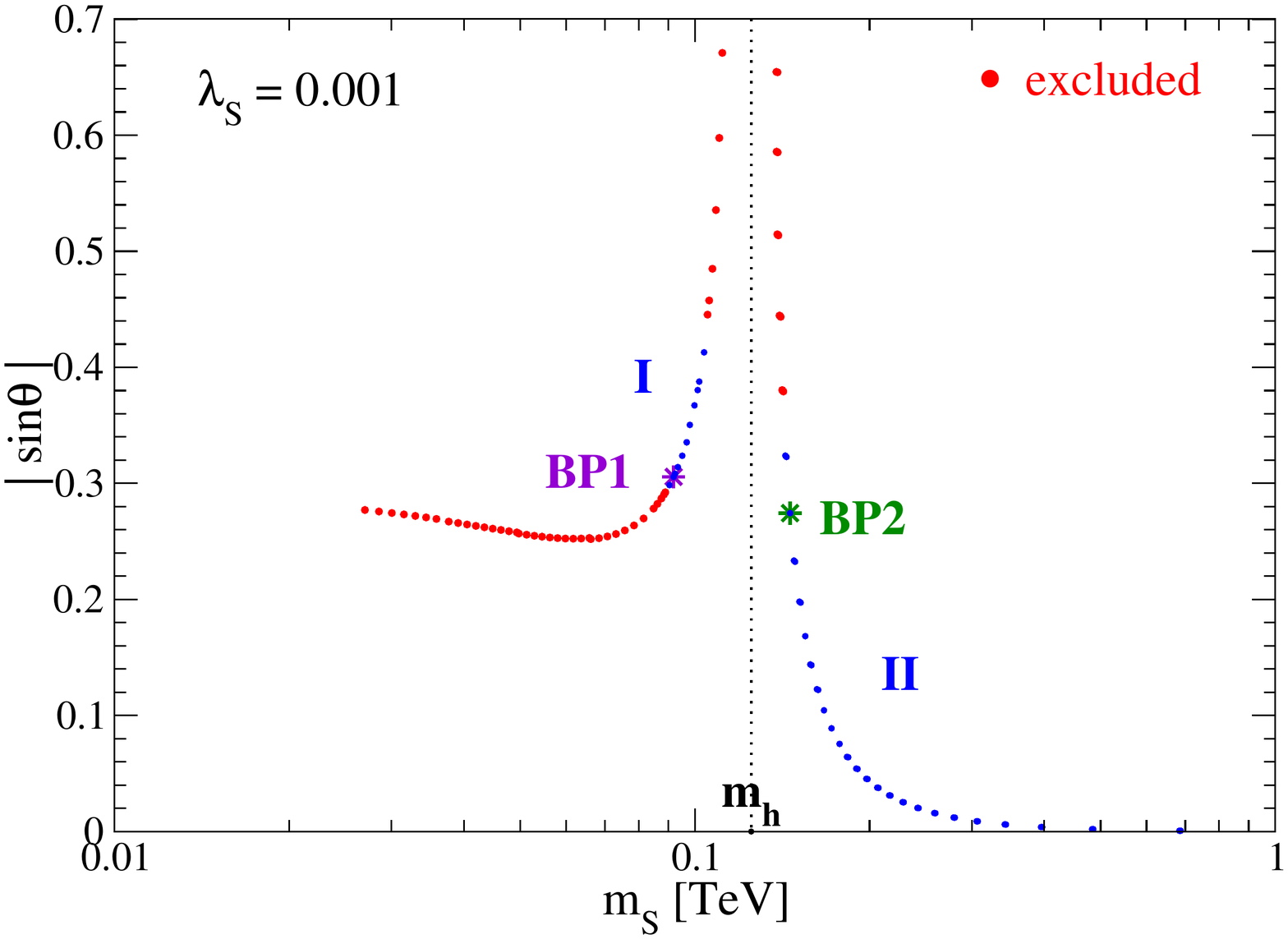}
\hspace{-0.2cm}
\includegraphics[width=3.2in]{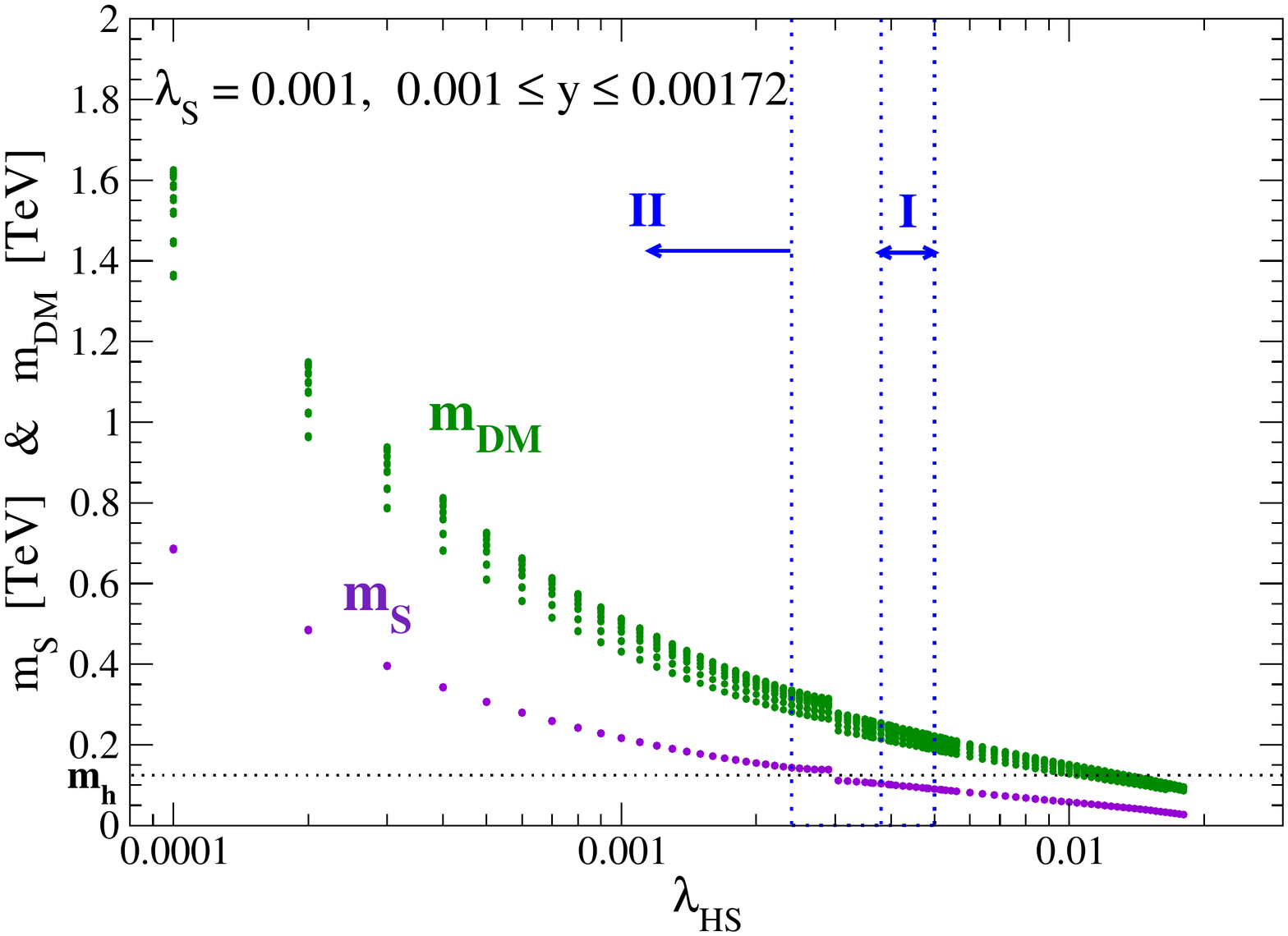}
\vspace{-5mm}
\caption{\footnotesize{Left: $|\sin\theta|$ vesus $m_S$ 
at $\lambda_S=0.001$, where we vary $\lambda_{HS}$
between  $0.0001$ and $0.018$ and 
$y$ between  $0.001$ and $0.00172$.
There are two branches; $m_S <  m_h$ and $m_S >  m_h$,
and on each branch there exist a (blue) region ( I and II)  that 
is allowed by LHC. Two benchmark points we consider 
are marked by BP1 (purple) and BP2 (green).
Right:
 $m_S$ (purple)  and $m_\text{DM}$ (green)  vesus $\lambda_{HS}$
at $\lambda_S=0.001$.}}
\label{sint-ms-mDM}
\end{center}
\end{figure}
The $h-S$ mixing  is constrained by LHC data 
(see e.g. refs. \cite{Falkowski:2015iwa,Robens:2016xkb}
and references thererin).
In the left panel of figure \ref{sint-ms-mDM} we plot 
$|\sin\theta|$ versus $m_S$ and in the right panel
$m_S$ (purple)  and $m_\text{DM}$ (green)  versus $\lambda_{HS}$
both at $\lambda_S=0.001$.
We vary $\lambda_{HS}$
between  $0.0001$ and $0.018$ and 
$y$ between  $0.001$ and $0.00172$.
(Why we consider $y$ in this interval will be  explained in Section
\ref{hQCD-PT}.)
As we see from the left panel, there are two branches; 
$m_S <  m_h$ and $m_S >  m_h$,
and on each branch there exist a (blue) region, I and II, that 
is allowed by LHC. 
The band of $m_\text{DM}$ in the right panel can be seen, because
it  sensitively depends on $y$, while $m_S$
is insensitive against $y$.
From each allowed region we choose a representative point,
BP1 and BP2, to get an orientation in the parameter space,
especially when discussing the GW spectrum  later on:
\be
\mbox{BP1} & :& \lambda_S = 0.001\, , ~\lambda_{HS} = 0.00485\, ,
~y= 0.00172\,,~\lambda_H=0.1238\,,~\Lambda_H= 4.322\,\mbox{TeV}\,,\nn\\
~\mbox{BP2} &:& \lambda_S = 0.001\, , 
~\lambda_{HS} = 0.00230\, ,
~y= 0.00170\,, ~\lambda_H=0.1325\,,
~\Lambda_H= 6.606\,\mbox{TeV}\,.\label{BP}
\ee

\subsection{Dark Matter}

Due to the vector-like flavor symmetry
(i.e. $SU(3)_V$ or its subgroup), the dark mesons are good DM candidates.
As we see from figure \ref{sint-ms-mDM}, the mass of the real singlet $m_S$ is
 smaller than the DM mass $m_{\mathrm{DM}}$,
so that the DM can annihilate into two $S$s in principle.
However, this annihilation cross section is  negligibly small
because  it is $\propto y^4 \lsim 10^{-11}$ in the parameter space
of interest.
To explain the observed value for the relic DM abundance
in this circumstance, we assume a
 hierarchy in the Yukawa couplings: $y_1=y_2<y_3$
(which breaks  $SU(3)_V$ down to $SU(2)_V\times U(1)$
explicitly), where $y_3$ should not differ very much from $y_2$ \cite{Ametani:2015jla}.
Under this assumption,
the dark mesons fall into three categories, $\tilde{\pi}=\left\{ \tilde\pi^\pm,~\tilde{ \pi}^0 \right\}, 
\tilde{K} =\left\{\tilde{K}^\pm,~\tilde{K}^0, ~\bar{\tilde{K}}^0 \right\}$
and $\tilde{\eta}$.
Here the dark mesons are given like the real-world mesons:
 \begin{align}
\tilde{\pi}^\pm &\equiv (\phi_1\mp i\phi_2)/\sqrt{2}~,~~
\tilde{ \pi}^0 \equiv \phi_3~, ~\nn\\
 \tilde{K}^\pm & \equiv (\phi_4\mp i  \phi_5)/\sqrt{2}~,
 ~~\tilde{K}^0(\bar{\tilde{K}}^0) \equiv
  (\phi_6+(-) i \phi_7)/\sqrt{2}~,~~ \tilde{\eta}^8 \equiv \phi_8\,,
  \label{mesons}
  \end{align}
where $\tilde{\eta}^8$ will mix with $\tilde{\eta}^0$ to form the mass eigenstates
$\tilde{\eta}$ and $\tilde{\eta}'$.
The states in the same category have the same mass, 
$m_{\tilde{\pi}^0}=m_{\tilde{\pi}^\pm} ( \equiv  m_{\tilde{\pi}}) $ and 
$m_{\tilde{K}^\pm}=m_{\tilde{K}^0}=m_{\bar{\tilde{K}}^0} ( \equiv m_{\tilde{K}} )$,
with $m_{\tilde{\pi}} < m_{\tilde{K}} < m_{\tilde{\eta}}$,
where the differences among 
$m_{\tilde{\pi}}$, $m_{\tilde{K}}$ and $m_{\tilde{\eta}}$
are small because of the small difference between $y_1=y_2$ and $y_3$.
The heavier state $\tilde{\eta}$ is an unstable NG boson which can mainly decay into two $\gamma$s.
On the other hand, the $\tilde{\pi}$ and $\tilde{K}$ are stable
due to the $SU(2)_V$ flavor symmetry
and become the DM.
Since the mass difference among
 $\tilde{\pi}, \tilde{K}$ and $\tilde{\eta}$ are small,
the DM annihilation into a pair of heavier DMs and/or $\tilde \eta$s, which are kinematically forbidden at zero temperature,
can become operative. 
In ref. \cite{Ametani:2015jla} it has been shown that the inverse conversion
$\tilde \pi\tilde\pi, \tilde K\tilde K\to \tilde\eta\tilde\eta\to \gamma\gamma\gamma\gamma$
can play a significant role 
to make the DM relic abundance realistic. 
This mechanism works only if the $SU(3)_V$ falvor symmetry is broken
into its subgroup.
\footnote{
In 
 the model considered in
 ref. \cite{Aoki:2017aws}  the   flavor group $SU(3)_V$ is unbroken and
 the hidden fermions have no coupling with the $U(1)_Y$ gauge boson.}

\begin{figure}[t]
\begin{center}
\includegraphics[width=3.7in]{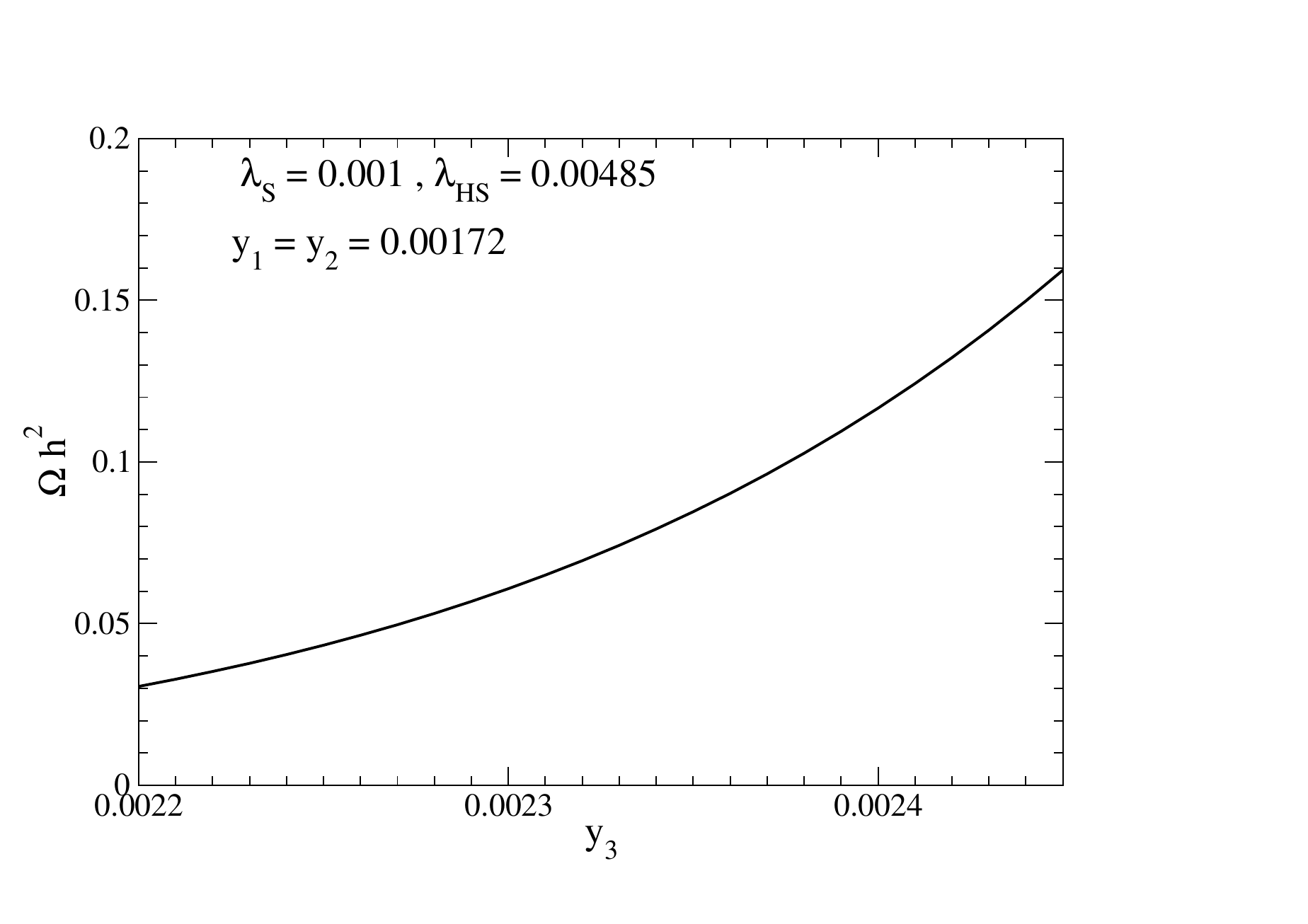}
\vspace{-5mm}
\caption{\footnotesize{The total DM relic abundance $\Omega h^2$ 
 with the $U(1)_Y$ hypercharge $Q=1/3$
vesus  $y_3$,
where we have fixed other parameters 
at  the benchmark point BP1 defined in eq. (\ref{BP}).}}
\label{y3-OM}
\end{center}
\end{figure}

In figure \ref{y3-OM} we show the total DM relic abundance
$\Omega h^2=\Omega_\pi h^2+\Omega_K h^2$
with the $U(1)_Y$ hypercharge $Q=1/3$
as a function of $y_3$, where the other parameters are
 chosen for  the benchmark point BP1 defined in eq. (\ref{BP}),
and $h$ is 
the dimensionless Hubble parameter.
We see from this figure that 
the DM relic abundance at $y_3\simeq 0.0024$ 
can coincide with  the experimentally observed value \cite{Aghanim:2018eyx}.
\footnote{Though the portal coupling 
$\lambda_{HS}$ is very small $\sim 10^{-3}$,
the singlet scalar $S$ decays via the $h-S$ mixing into the SM particles 
 much before Big Bang Nucleosynthesis (BBN):
$\Gamma(S\to \mbox{SM particles})/H \simeq \sin^2\theta\,
\Gamma(h\to \mbox{SM particles})/H\sim
\sin^2\theta\times 10^{22}$ at $T=1$ MeV.}

\section{Chiral Phase Transition and 
Bounce Solution}\label{hQCD-PT}

\subsection{Effective potential and chiral phase transition}\label{hQCD-PT-A}
\begin{figure}[t]
\begin{center}
\includegraphics[width=3.7in]{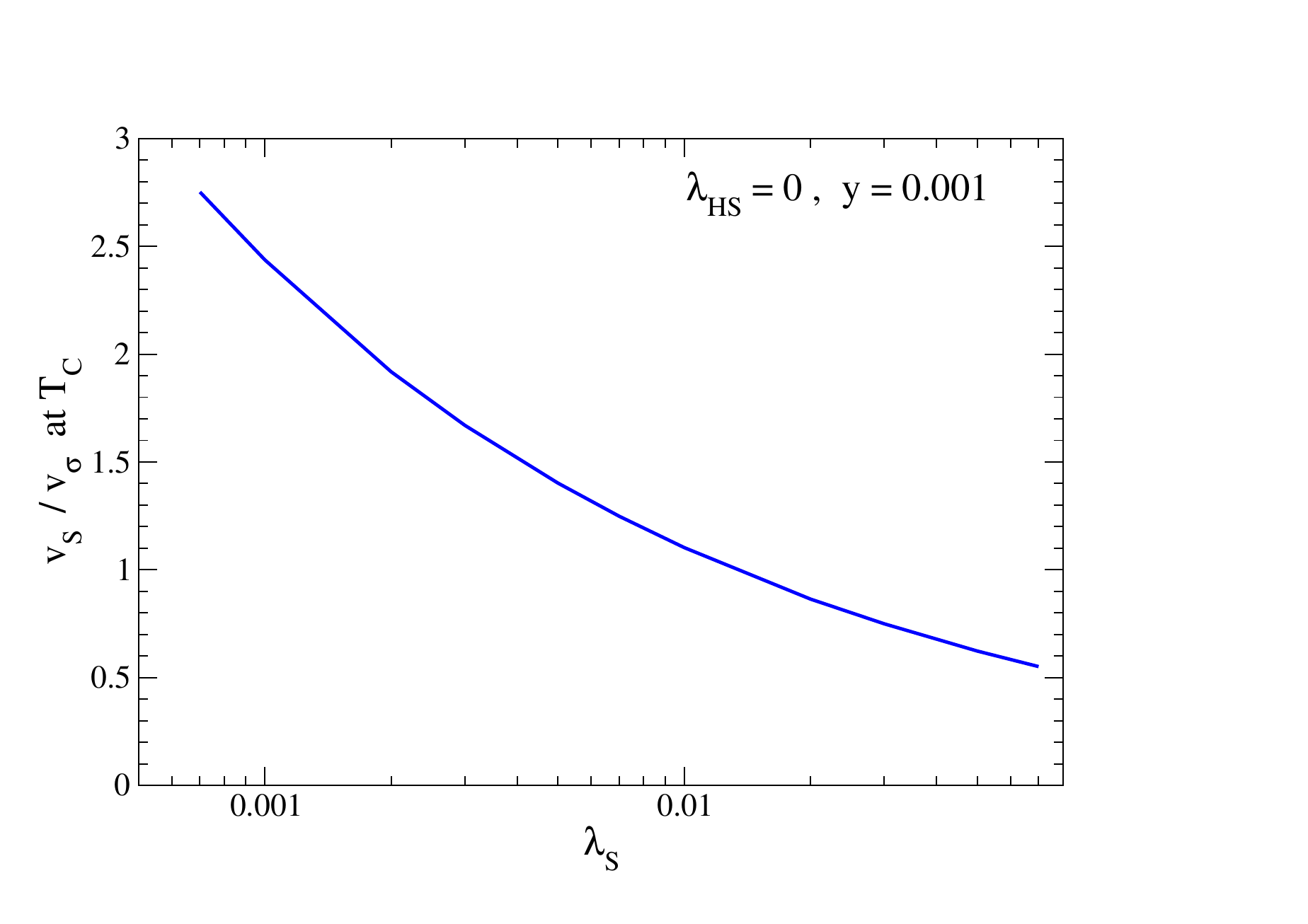}
\vspace{-5mm}
\caption{\footnotesize{$v_S/v_\sigma$ at the critical temperature
$T_C$ against $\lambda_S$. 
We see that the smaller $\lambda_S$ is, the  larger 
the ratio $v_S/v_\sigma$ becomes, which means more
deviation from the pure NJL (i.e. without the singlet scalar $S$).}
}
\label{rs-vsvsigma}
\end{center}
\end{figure}

\begin{figure}[t]
\begin{center}
\includegraphics[width=3.35in]{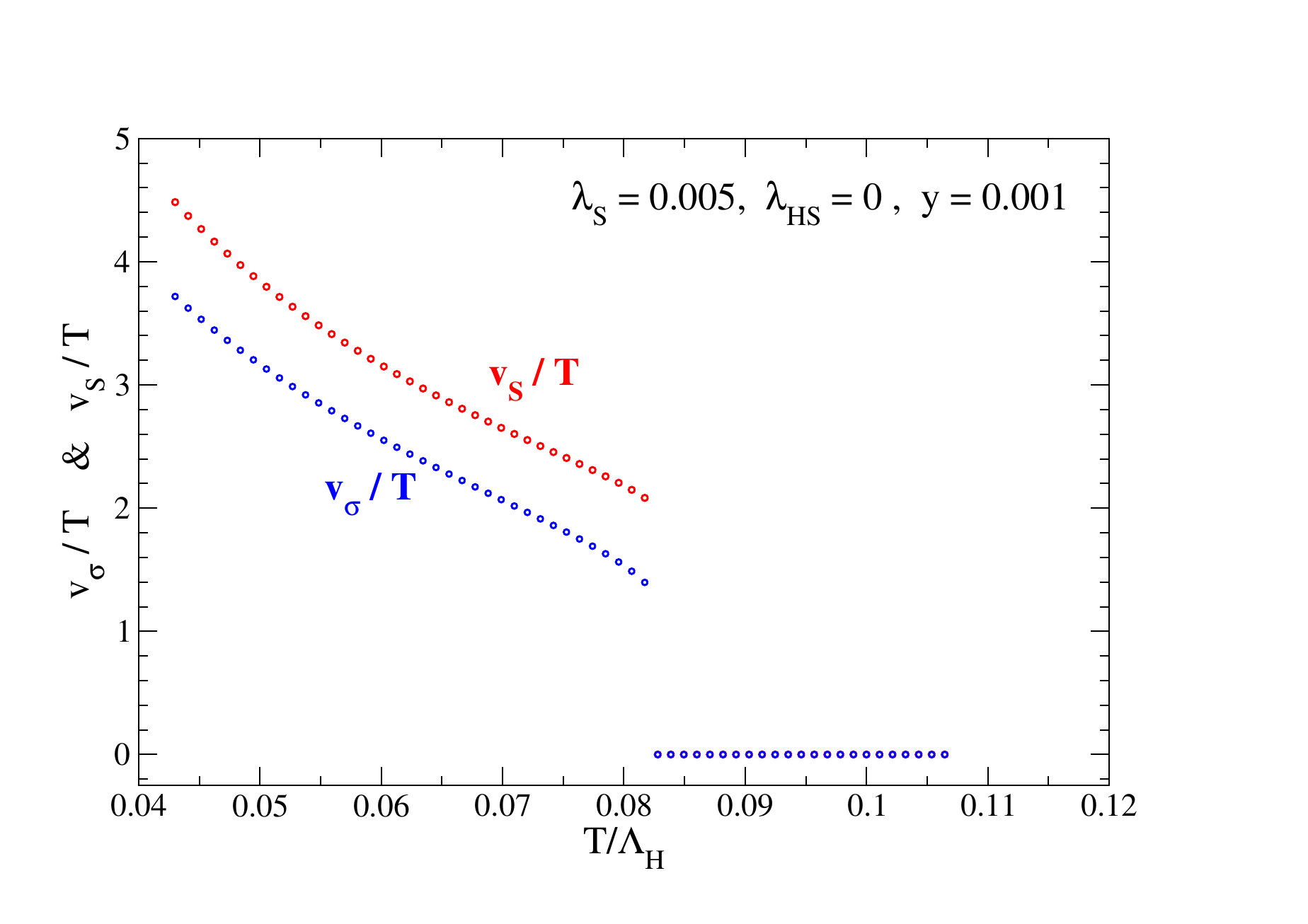}
\hspace{-1cm}
\includegraphics[width=3.35in]{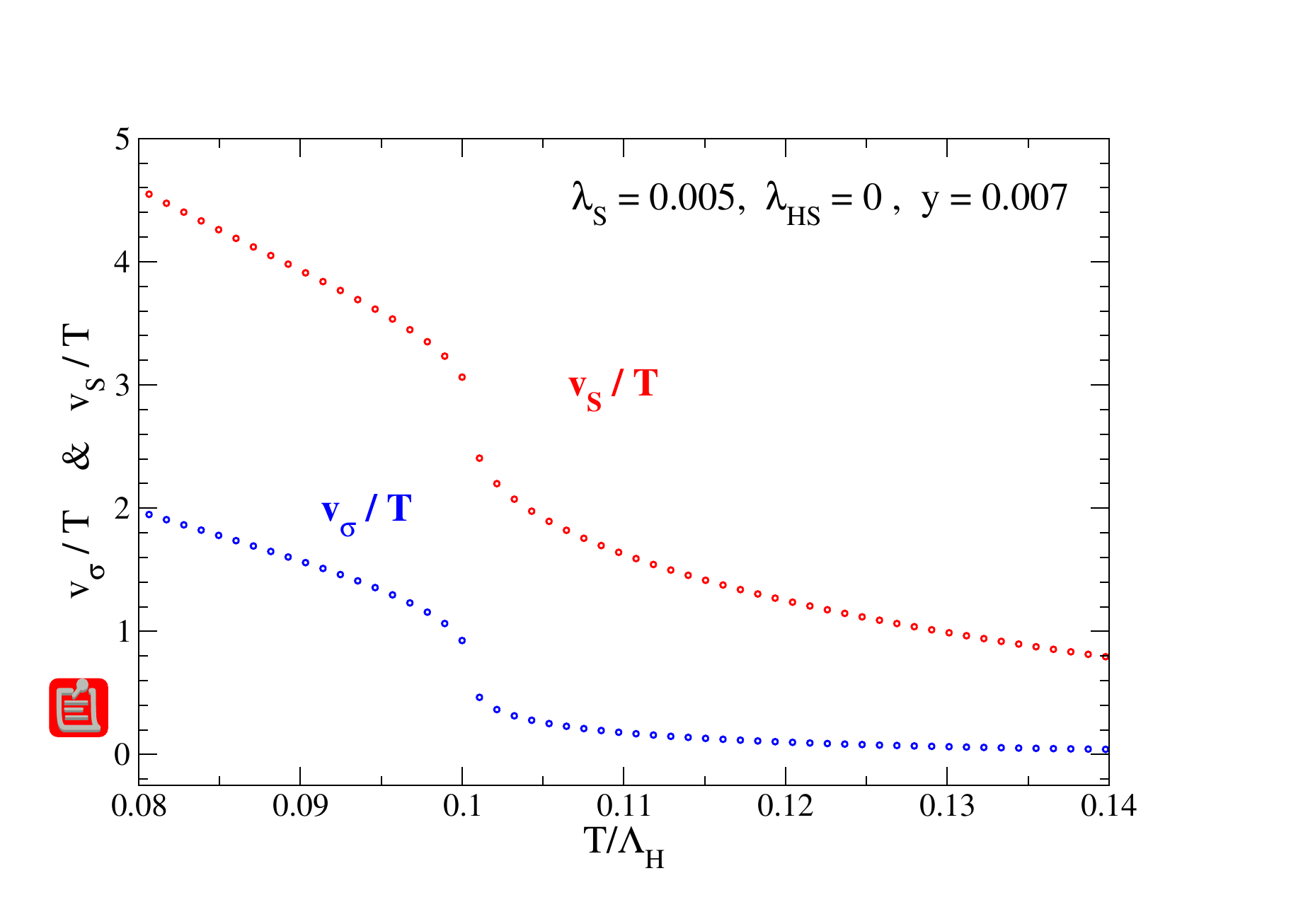}\\
\vspace{-3.5mm}
\includegraphics[width=3.35in]{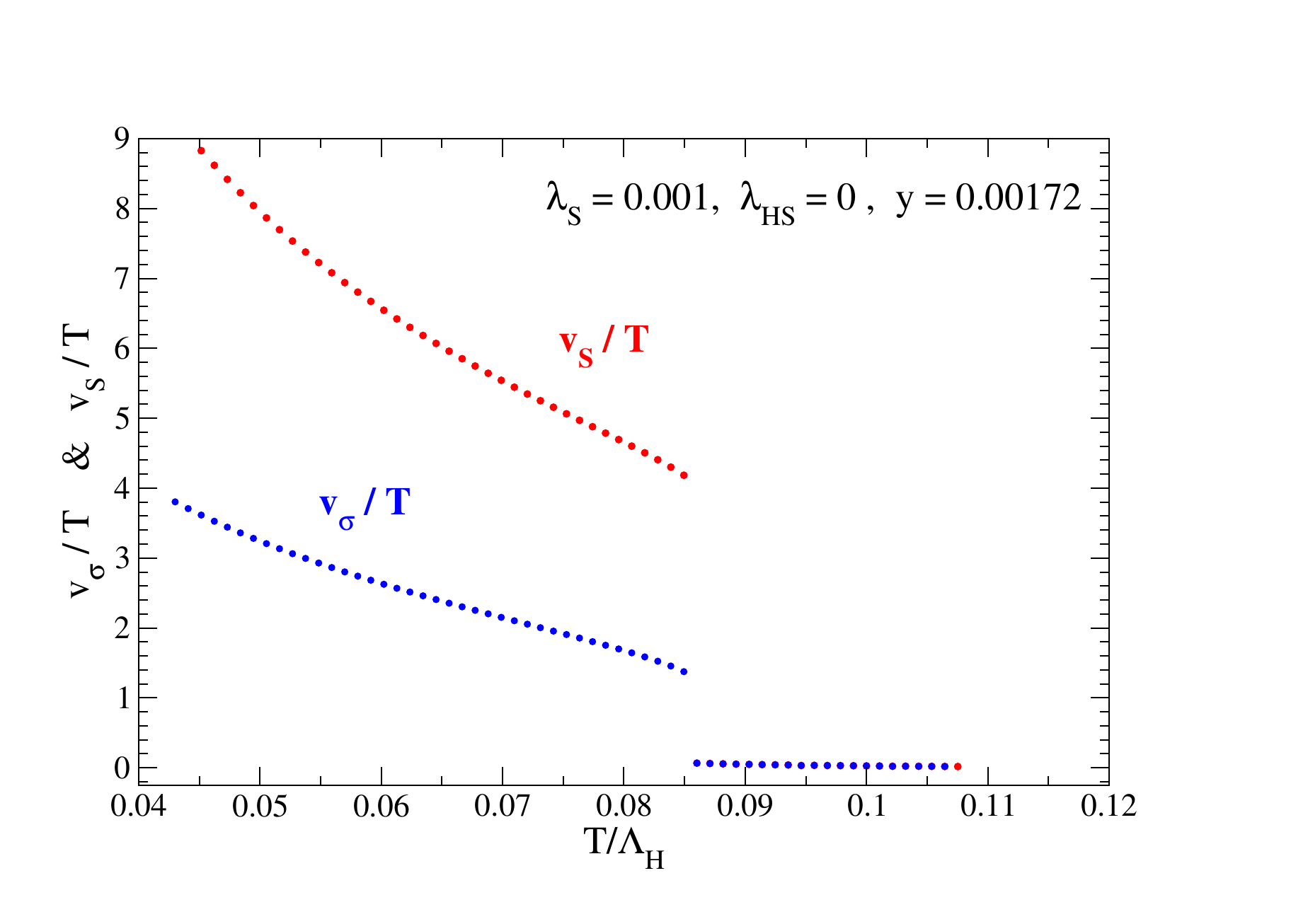}
\hspace{-1cm}
\includegraphics[width=3.35in]{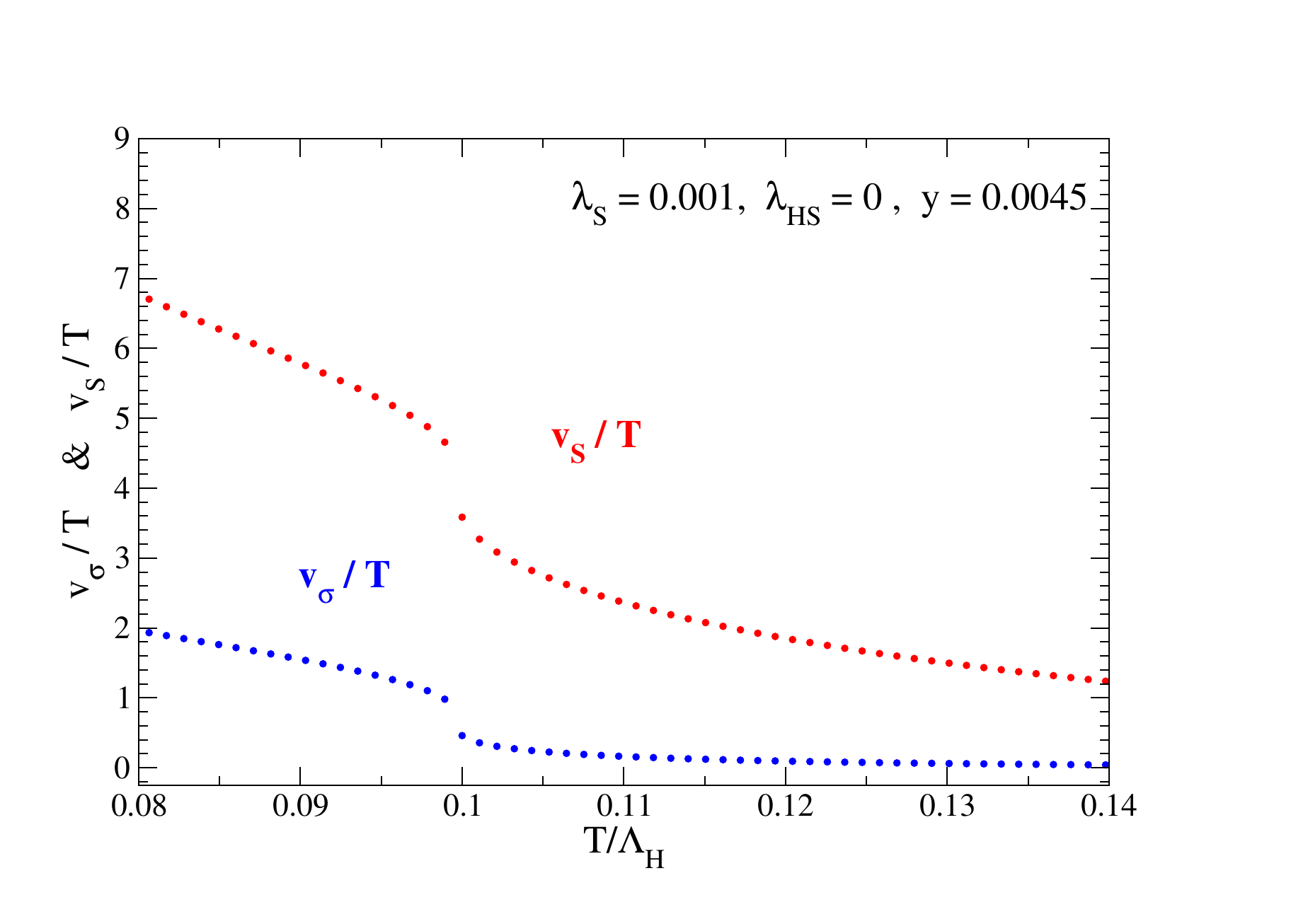}
\vspace{-5mm}
\caption{\footnotesize{Upper (lower) left:
$v_\sigma/T$ (blue) and $v_S/T$ (red)  against  $T/\Lambda_H$
for $\lambda_S=0.005~(0.001), \lambda_{HS}=0, y=0.001~(0.00172)$,
showing a strong first-order PT.
Upper (lower) right: The same with $y=0.007~(0.0045)$, showing a 
transition of cross-over type.}}
\label{PT}
\end{center}
\end{figure}

The EW  and  chiral PTs in our model (\ref{eq:LH})
have been studied in refs. \cite{Holthausen:2013ota,Aoki:2017aws}
in some detail.
For a phenomenologically viable region of the parameter space,
the EW PT occurs -- with decreasing temperature -- after the chiral PT
takes place in a hidden sector.
Therefore, the VEV of $h$ vanishes during the chiral PT, so that 
we  set $\left<h\right>=0$   in investigating the chiral PT.
Accordingly, we analyze the following scalar potential at finite temperature:
\begin{align}
\label{VEFF}
   V_{\text{EFF}} (S,\sigma,T) =  V^{h\rightarrow0}_{\text{SM}+S}(S)+V_{\text{NJL}} (S,\sigma)+V_{\text{CW}} (S)
   +V_{\text{FTB}}(S,T)   +V_{\text{FTF}}(S,\sigma,T)\,,
\end{align}
where $V_{\text{SM}+S}$ and $V_{\text{NJL}}(S,\sigma)$ are given in eqs. (\ref{eq:VSM}) and  (\ref{eq:Vnjl}), respectively, 
\begin{align}
V_{\text{CW}} (S)&= \frac{m^4_S(S)}{64\pi^2}
\left[ \ln (S^2/\langle S\rangle ^2)-1/2 \right]\,,\\
V_{\text{FTB}}(S,T)&= \frac{T^4}{2\pi^2}
\left[J_{B}(M_S(S)/T)-
J_B(|\lambda_S/4-\lambda_{HS}/6|^{1/2})\right]\,, \\
V_{\text{FTF}}(S,\sigma,T)&=  -6n_c\frac{T^4}{\pi^2}
\left[J_{F}(M(S,\sigma)/T)-J_F(0)\right]\,,
\label{VFTF}\\
M^2 _S &= m_S ^2(S) +\left(\frac{\lambda_S}{4}-\frac{\lambda_{HS}}{6}\right)T^2~\mbox{with}~m_S ^2(S)=3\lambda_S S^2\,,
\label{thermalM}
\end{align}
and $M(S,\sigma)$ is given in (\ref{M}).
The thermal functions are
\begin{align}
\label{thermalFB}
  J_{F/B}(u) & =\int^{\infty} _{0} dx x^2 \ln 
  \left( 1\pm  e^{-\sqrt{x^2+u^2}}\right)\nn\\
 &=\sum_{j=1}^\infty (\mp1)^{(1+j)}(u^2/j^2)
K_2(j \,u)\,,
\end{align}
where $K_2(j\,u)$ is the modified Bessel function of the 
second kind of order two, and we will truncate the sum at $j=10$.
\footnote{
The error  of 
 the approximate
function $J_{F(B)}(u;j_{\rm max} )$
 with the truncation at $j = j_{\rm max}$
is $\Delta J_{F(B)}=\left|\, J_{F(B)}(u)-J_{F(B)}(u;j_{\rm max} )\,
\right|$. 
For $j_{\rm max} =10$, $\Delta J_{F}
< 5\times 10^{-5}$ and $\Delta J_{B}
< 3\times 10^{-4}$ are satisfied.
So, with  $j_{\rm max}=10$,
the error in 
$\Delta (V_{\rm EFF}/T^4) $ is less than $10^{-4}$.
Since we are interested in the behavior of $V_{\rm EFF}$ near
the critical temperature $T_C$,
we obtain $\Delta (V_{\rm EFF}/\Lambda_H^4) 
< 10^{-8}$ because $T_C/\Lambda_H \sim 0.1$.
This accuracy is sufficient for our purpose.
}
In $V_{\text{FTB}}(S,T)$ and 
$V_{\text{FTF}}(S,\sigma,T)$ we have subtracted
the (temperature-dependent) constant terms such that 
$V_{\text{FTB}}(0,T)=V_{\text{FTF}}(0,0,T)=0$. We first note that 
the role of the singlet scalar $S$ becomes more important
for smaller $\lambda_S$.
To see this, we will consider $V_{\text{NJL}} (S,\sigma)$
for small $y$:
\al{
V_{\text{NJL}} (S,\sigma)
& =
V_{\text{NJL}} (0,\sigma)- \frac{3 n_c\Lambda_H^2\sigma}{4 \pi^2} y S
+O((y S)^2)\,.}
Since, neglecting the portal coupling
$\lambda_{HS}$, the scalar potential $V_{\text{SM}+S}$ becomes $\lambda_H (H^\dag H)^2+(1/4)\lambda_{S}S^4$, 
 we  find
\al{
v_S^3 &=\langle S\rangle^3
\simeq  \frac{3 n_c\Lambda_H^2 v_\sigma}{4 \pi^2}
\left( \frac{y}{\lambda_S}\right)\,.}
Therefore,
the deviation 
from the pure NJL model (i.e. without the singlet scalar $S$)
is larger for smaller $\lambda_S$ and larger $y$.
The above feature remains at finite temperature, as we can see from 
figure \ref{rs-vsvsigma}, where 
 we show $v_S/v_\sigma$ against $\lambda_S$ at the critical temperature 
$T_C$ for $y=0.001\,, \lambda_{HS}=0 $.\footnote{The absolute scale of
the hidden sector (i.e. $\Lambda_H$)  can be anything  when the hidden sector
has no coupling with the SM sector. This happens, for instance, when 
$\lambda_{HS}$ is set equal to zero.
Nevertheless,  dimensionless quantities have their meaning.
As we see from eqs. (\ref{eq:VSM}) and (\ref{VEFF})
the $\lambda_{HS}$ dependence of the 
effective potential is very small if $\langle h\rangle=0$: 
It enters only in the thermal mass of $S$ as one can see in eq. (\ref{thermalM}).}
In figure \ref{PT} we plot $v_\sigma/T$ (blue) and 
$v_S/T$ (red) as a function of $T/\Lambda_H$
for $\lambda_S=0.005, \lambda_{HS}=0 $ and $y=0.001$ (upper left panel) and $0.007$ (upper right panel), respectively.
We see that the chiral PT is no longer
a strong first-order PT at $y=0.007$.
The lower panels show the case for
$\lambda_S=0.001, \lambda_{HS}=0 $ and $y=0.00172$ (left) and $0.0045$ (right).

\subsection{Bounce solution}\label{Bounce }

One of the main quantities 
in discussing the stochastic GW background 
produced by a first-order PT in the expanding Universe
 is the duration time of the first-order PT, 
$\tau_\text{pt}=\beta^{-1}$,  
which should be  compared  with the inverse rate of the  expansion $H^{-1}$
\cite{Witten:1984rs,Kosowsky:1991ua,Kosowsky:1992rz,Kosowsky:1992vn}.
In fact the GW energy density 
increases -- depending on the nature of its source -- linearly or even
quadratically with $(\beta/H)^{-1}$ (see eqs. (\ref{omephi}), (\ref{omesw})
and (\ref{ometurb})), 
while its peak frequency increases linearly 
with $\beta/H$ (see eqs. (\ref{fpeakphi}), (\ref{fpeak})
and (\ref{fpeakturb})). 
In ref. \cite{Helmboldt:2019pan} it has been found that 
$\beta/H$ is of order $10^{4}$
in the pure NJL model,
so that  $\Omega_\text{GW}$, the spectral GW  energy density 
normalized to the critical energy density, is considerably suppressed.
Therefore, we consider here a parameter space in which
the deviation from the pure NJL model is large.
From the discussion of Section \ref{hQCD-PT-A} we can infer that
the area with  small $\lambda_S$ and  large $y$ is
an optimistic parameter space, where
$y$ should not be too large for a strong first-order chiral PT 
to be realized.
It turns out that $\lambda_S\sim O(10^{-3})$ is an optimistic
magnitude for  $\lambda_S$, and in the following discussions we
concentrate on the parameter space with  $\lambda_S=0.001$.

To obtain $\beta/H$ we have to compute the value of the corresponding $O(3)$
symmetric  Euclidean action $S_3$ \cite{Linde:1981zj}.
The mean field $\sigma$, introduced as an auxiliary field at the tree level
in the mean-field approximation, is a driving force for the chiral PT.
As it has been discussed in Section \ref{Mass spectrum} the mean field $\sigma$ 
can be promoted to a propagating quantum field at one loop,
which also applies at finite temperature.
The kinetic term for $\sigma$  at finite temperature has been correctly
computed in ref. \cite{Helmboldt:2019pan}. Quoting the result of ref.
 \cite{Helmboldt:2019pan},
the  $O(3)$ symmetric  action $S_3$ can be written as
 \al{
  \label{actionS}
S_3(T)=4 \pi \int d r r^2\left[\frac{Z^{-1} _{\sigma}(S,\sigma,T)}{2}
\left(\frac{d \sigma}{dr}\right)^2+\frac{1}{2}
\left(\frac{d S}{dr}\right)^2+
V_{\mathrm{EFF}}(S,\sigma,T) \right]\,,
}
where $r$ is the radial coordinate of the 3-dimensional space,
and $V_{\mathrm{EFF}}(S,\sigma,T)$ is given in eq. (\ref{VEFF}).
$Z_\sigma(S,\sigma,T)$ is the wave function renormalization
``constant'' at finite temperature \cite{Helmboldt:2019pan}:
\be
& & Z^{-1} _{\sigma}(\sigma,S,T)=
  \frac{n_c n_f}{2\pi^2} \left[1-\frac{G_D}{4G^2}\sigma\right]^2
 \nn\\
 &\times & \left\{-\frac{\Lambda_H^2}{4(\Lambda_H^2+M^2)}+
 \frac{1}{4}\ln(1+\Lambda_H^2/M^2)
  +\frac{\Lambda_H^4}{8(\Lambda_H^2+M^2)^2}
 -  \frac{\Lambda_H^4(\Lambda_H^2+3M^2)}{6(\Lambda_H^2+M^2)^3}\right.\nn\\
&  & +\int_0^{\infty}dx  x^2\left[- \frac{1+e^{\omega/T}(1+
\omega/T)}{(1+e^{\omega/T})^2
 (\omega/T)^3}
  - (M/T)^2 \frac{3+e^{\omega/T}(6+3\omega/T-
 (\omega/T)^2)}{4(1+e^{\omega/T})^3
 (\omega/T)^5}
 \right.\nn\\
 &&\left. - (M/T)^2 \frac{e^{2\omega/T}(3+3\omega/T+
 (\omega/T)^2)}{4(1+e^{\omega/T})^3
 (\omega/T)^5}
 \right.\nn\\
 & &  \left. +(M/T)^4 \frac{15+e^{\omega/T}(45+15\omega/T-
 6(\omega/T)^2+(\omega/T)^3)
  +e^{2\omega/T}(45+30\omega/T-
 4(\omega/T)^2)}{6(1+e^{\omega/T})^4
 (\omega/T)^7}\right.\nn\\
 & &\left.\left.+ (M/T)^4 \frac{e^{3\omega/T}(15+15\omega/T
+6(\omega/T)^2)+(\omega/T)^3)}{6(1+e^{\omega/T})^4
 (\omega/T)^7}\,\right]\right\}\,,
 \label{Z}
\ee
where
\be
\omega/T &= &[x^2+(M/T)^2]^{1/2}\,,
\ee
and $M$ is given in eq. (\ref{M}).

\begin{figure}[t]
\begin{center}
\includegraphics[width=3.35in]{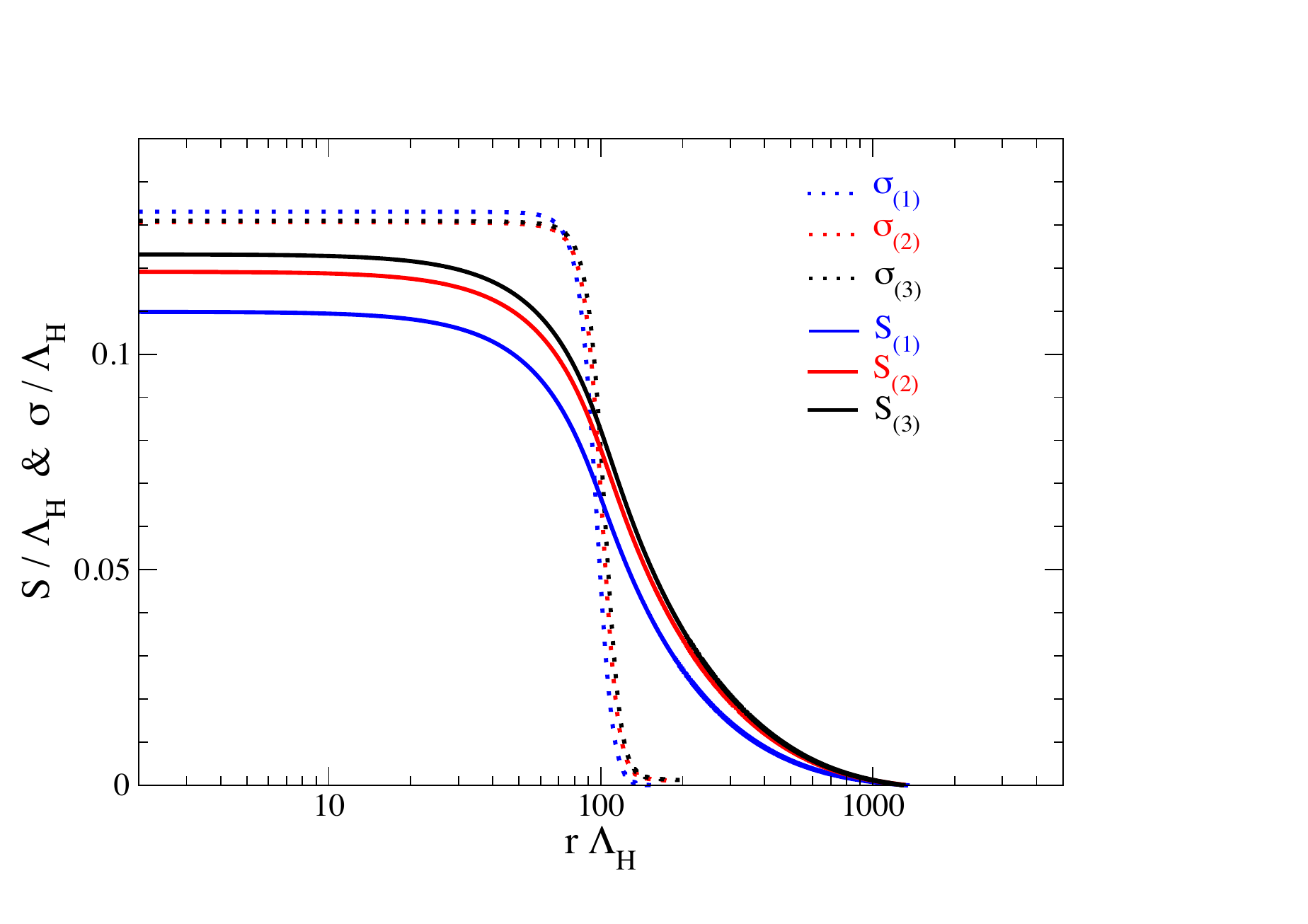}
\hspace{-1cm}
\includegraphics[width=3.35in]{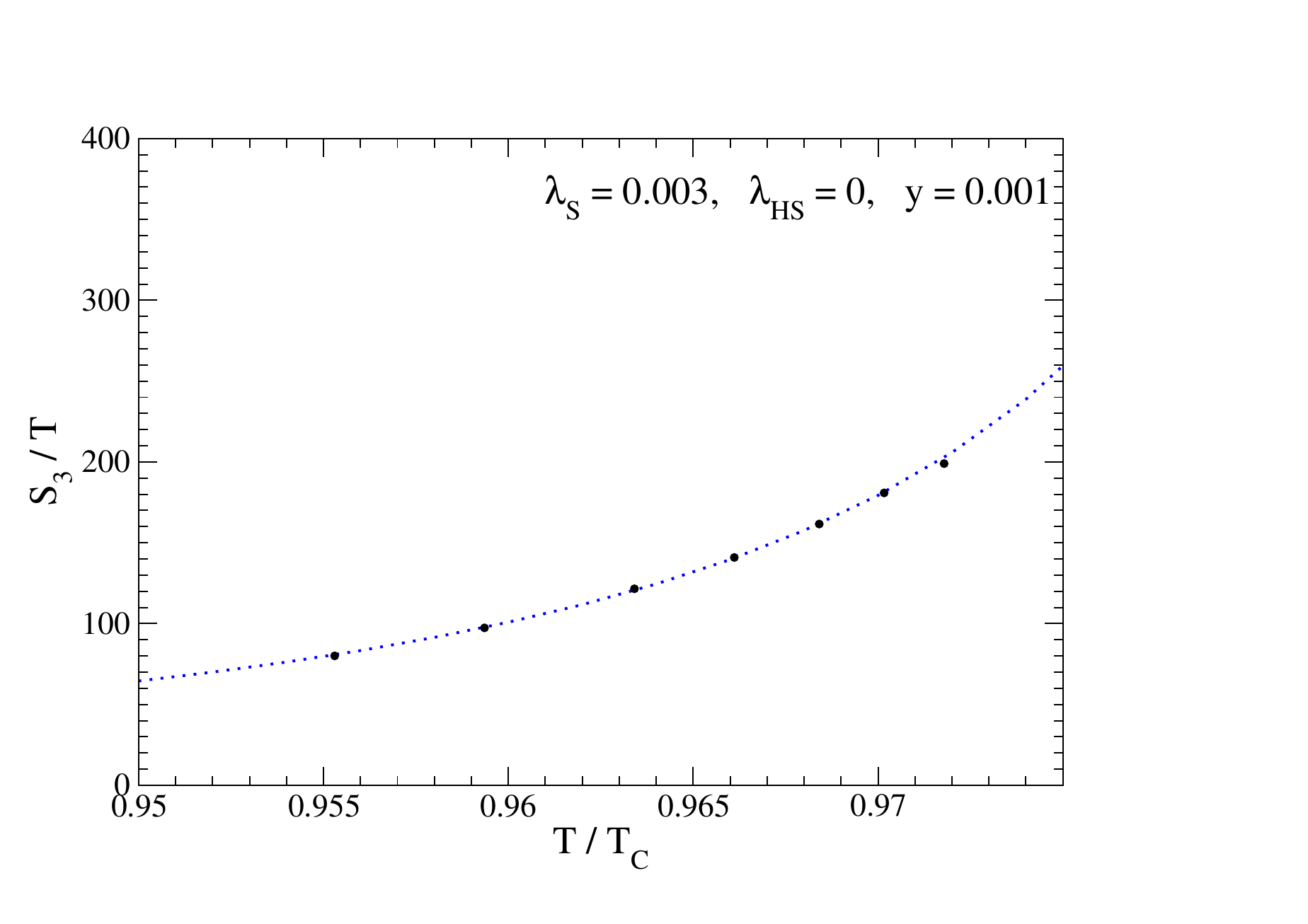}\\
\vspace{-3.5mm}
\includegraphics[width=3.35in]{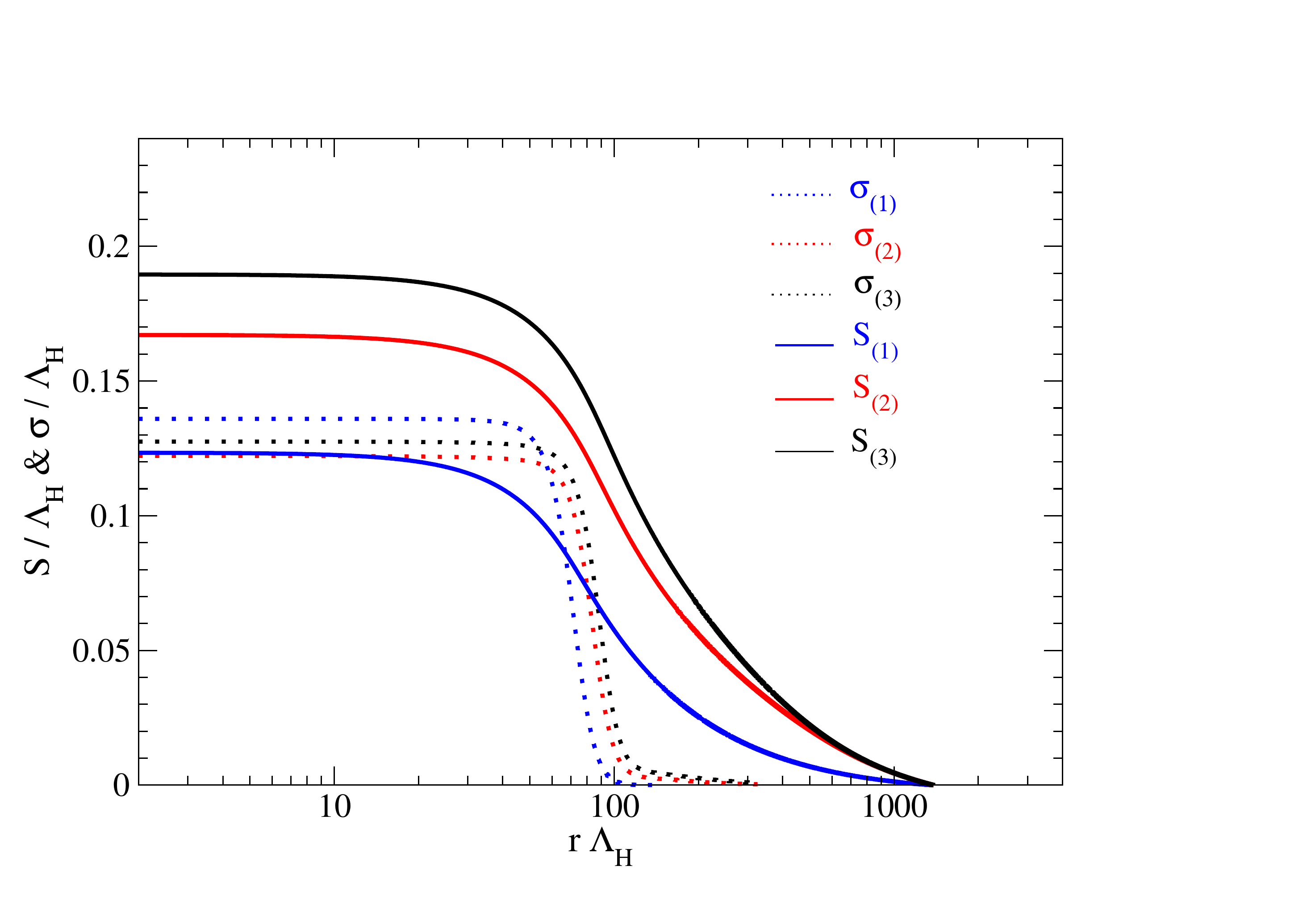}
\hspace{-1cm}
\includegraphics[width=3.35in]{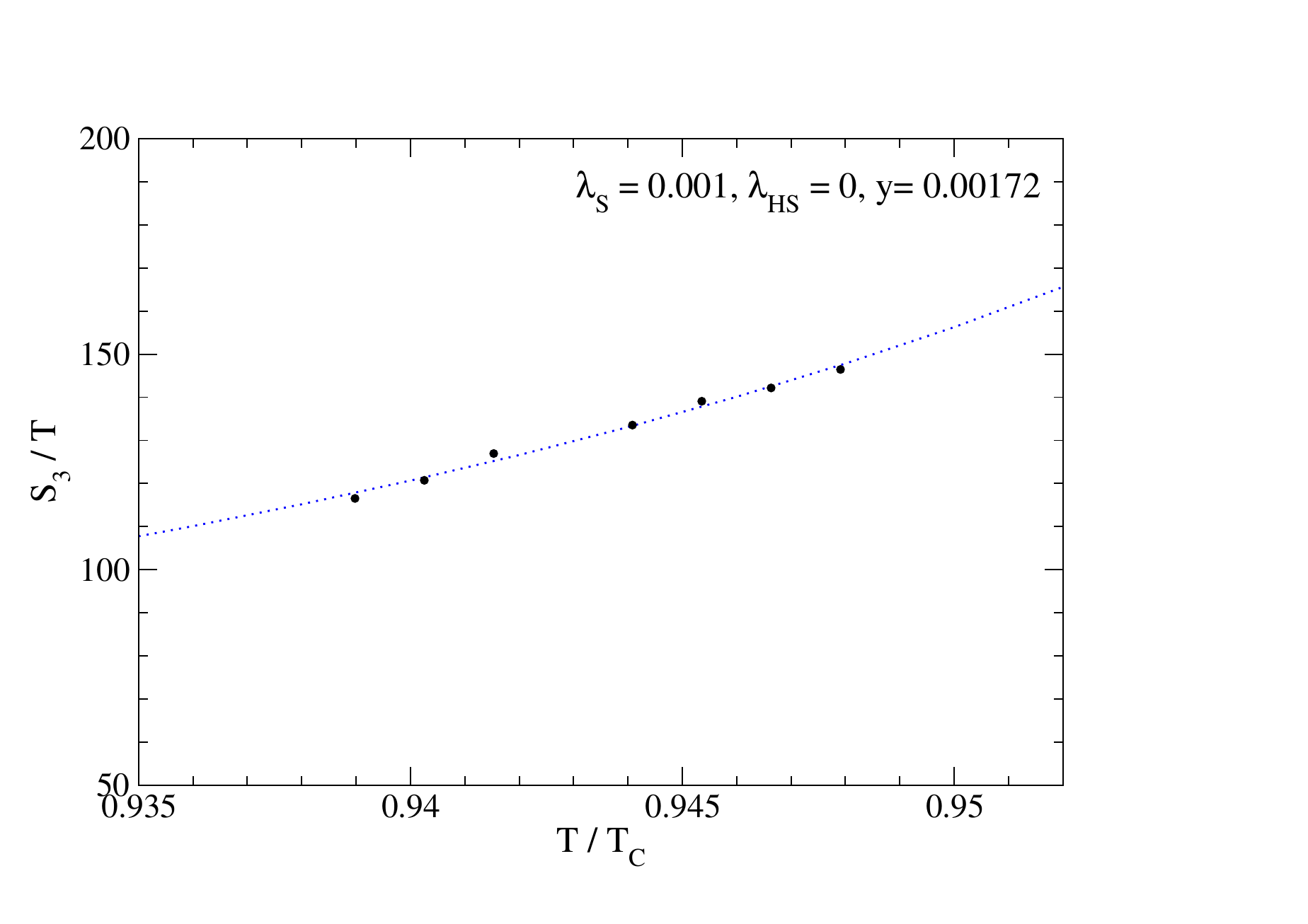}
\vspace{-5mm}
\caption{\footnotesize{Upper (lower) left:
 The bounce solutions  $\sigma_{(n)}(r)$ and  
 $S_{(n)}(r)$ for 
 $n=1\,(\mbox{blue}),\, 2\,(\mbox{red})$ and $3\,(\mbox{black})$,
 where we have used the parameters:
$\lambda_S=0.003\, (0.001)\,, \,\lambda_{HS}=0\,, \,y=0.001\,(0.00172)\,,
 T_C/\Lambda_H =0.0796\, (0.0843)\,, \,T/\Lambda_H =0.0769\, (0.0798)$.
Upper (lower) right:
 $S_3/T$ (in the second round) against $x=T/T_C$
with the same $\lambda_S\,,\lambda_{HS}$ and $y$
as in the upper (lower) left panel. 
The black points are obtained by applying  our iterative method,
while the blue dotted line is the fitting function defined
in eq. (\ref{fit}) with $b=0.1833\, (2.232)$ and $\gamma=1.961\,(1.418)$.}
 }
\label{bounce}
\end{center}
\end{figure}

The  equations of motion for the action (\ref{actionS}) read
\begin{align}
      \label{dfeqs}
  \frac{d^2 \sigma}{dr^2} + \frac{2}{r}\frac{d \sigma}{dr}
  +\frac{1}{2}\frac{\partial \ln Z^{-1}_{\sigma}(S,\sigma,T)}{\partial \sigma}\left(\frac{d \sigma}{dr}\right)^2 & =Z_{\sigma}(S,\sigma,T)\frac{\partial V_{\mathrm{EFF}}(S,\sigma,T) }{\partial \sigma},\\
   \label{dfeqS}
  \frac{d^2 S}{dr^2} + \frac{2}{r}\frac{d S}{dr} 
-\frac{1}{2}\frac{\partial  Z^{-1}_{\sigma}(S,\sigma,T)}{\partial S}\left(\frac{d \sigma}{dr}\right)^2& =\frac{\partial V_{\mathrm{EFF}}(S,\sigma,T) }{\partial S}\,,
\end{align}
and the boundary conditions for the bounce solution  are 
given by  \cite{Linde:1981zj}
\begin{align}
\label{boundary}
\left.\frac{d\sigma}{dr}\right|_{r=0}=0\,,~\left.\frac{dS}{dr}\right|_{r=0}=0\,,~\lim_{r\rightarrow \infty}\sigma(r)=0\,,
~\lim_{r\rightarrow \infty}S(r)=0\,.
\end{align}
The bounce solution describes a bubble, where
$r=0$ is the center of the  bubble, inside  of which the chiral symmetry
is broken.
The bubble however  has no sharp boundary,
but $\sigma(r)$ and $S(r)$ at  $r\simeq r_w$
drop sharply from a finite value to a small value 
(see figure \ref{bounce}) so that 
$r_w$ can be understood as  the position of the bubble wall:
We may say in a less precise way that
inside  of the wall the chiral symmetry is broken and in the outside
of the wall it is unbroken.
In the one-dimensional case, in which there exists only one scalar
degree of freedom as an order parameter,
we can obtain a bounce solution by using 
the so-called overshooting/undershooting method \cite{Apreda:2001us}. 
However,  this is an extremely cumbersome 
method in a multi-dimensional case,
because a set of certain initial conditions have to be simultaneously  fine tuned. 
An appropriate method is 
the path deformation method \cite{Wainwright:2011kj}.
But to minimize the problem associated with the complicated structure of
the wave function renormalization (\ref{Z}), 
we here use another iterative method,
which we will describe below.

One round of the calculation consists of two steps.
At the first step in the $n^\text{th}$ round,  
we solve the differential equation (\ref{dfeqs}) for $\sigma(r)$ with
 $S(r)=S_{(n-1)}(r)$, where $S_{(n-1)}(r)$ 
 is obtained in the  $(n-1)^\text{th}$ round. 
The solution is denoted by  $\sigma_{(n)}(r)$.
At the second step in the $n^\text{th}$ round,
we solve the differential equation (\ref{dfeqS}) for  $S(r)$ with $\sigma(r)=
\sigma_{(n)}(r)$ to obtain  $S_{(n)}(r)$.
Then, using $\sigma_{(n)}(r)$  and $S_{(n)}(r)$ we compute $S_3/T$ in 
the $n^\text{th}$ round and denote it by $(S_3/T)_{(n)}$.
Since  each step   is a one-dimensional problem, we 
apply the overshooting/undershooting method.
Of course, there is no mathematical warranty that  the iterative process
converges: It depends strongly on  
 $S_{(0)} $ that is needed to  carry out
  the first step in the first round, i.e., to obtain   $\sigma_{(1)}(r)$.
 Here  we assume that  $S_{(0)} $ is a function of $\sigma$
and choose  it as a straight line linking
 the origin of the field space ($\sigma=S=0$)
 and the position $(v_\sigma,v_S)$ of the minimum of
 $V_\text{EFF}(S,\sigma,T < T_C)$:
 \be
 S_{(0)}(\sigma) &=&
\frac{v_S}{v_\sigma} \sigma\,.
 \ee
In the upper (lower) left 
 panel of figure \ref{bounce}, we show  
 $\sigma_{(n)}(r)$ and  
 $S_{(n)}(r)$ with $n=1\, (\mbox{blue})$, $2\, (\mbox{red})$, $3\, (\mbox{black})$
  for $\lambda_S=0.003$ $(0.001)\,, \,\lambda_{HS}=0\,,
   \,y=0.001\,(0.00172)\,,
\, T_C/\Lambda_H =0.0796\,(0.0843)$, $T/\Lambda_H =0.0769\,(0.0798)$.
We find also
\be
 (S_3/T)_{(1)} &=& 144.4\, (153.5)\,,
  ~~~(S_3/T)_{(2)} = 140.9\, (140.0)\,,\nn
  \\ (S_3/T)_{(3)} &=& 141.5\, (144.7)\,,
  ~~~(S_3/T)_{(4)} = 140.8\, (142.2)\, .
  \ee
So, the convergence of the iterative process,
 described  above, is quite fast.
We have calculated $S_3/T$ for several values of $x=T/T_C$ 
in the second round and found
that $S_3/T$ for $x <1$ can be nicely fitted with  a simple function
\cite{Hogan:1984hx,Helmboldt:2019pan}
\be
\frac{S_3}{T}(x) &=& b\left(1-x\right)^{-\gamma}\,.
\label{fit}
\ee
This is shown in the upper (lower)  right panel of figure \ref{bounce} for 
$\lambda_S$=0.003 (0.001)\,, $\lambda_{HS}=$0\,, y=0.001 (0.00172).
The blue dotted line is the function (\ref{fit}) 
with $b=$0.1833 (2.232) and $\gamma$=1.961 (1.418).

\begin{figure}[t]
\begin{center}
\includegraphics[width=3.1in]{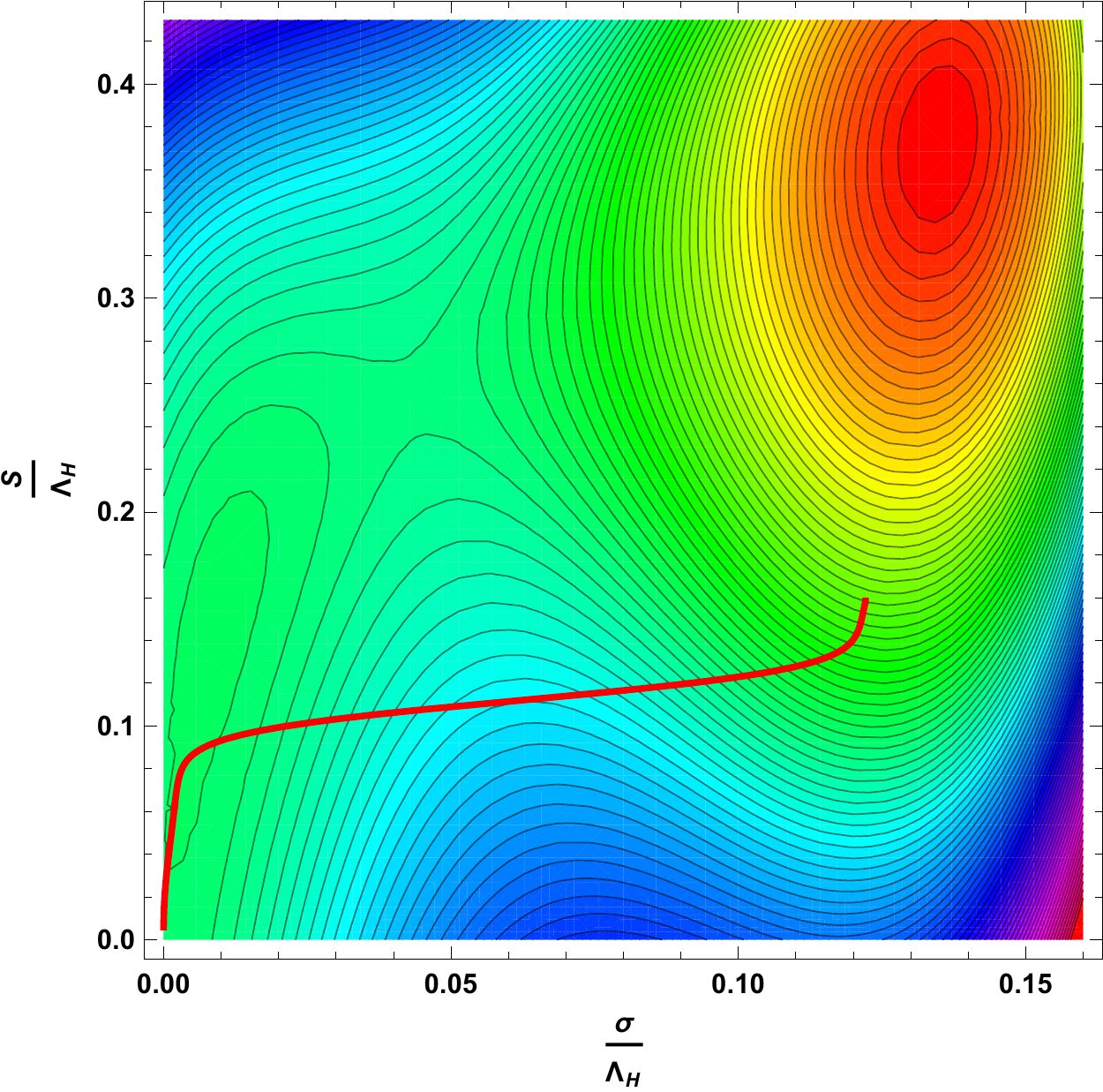}
\vspace{-2mm}
\caption{\footnotesize{The contour plot of the effective potential $V_\text{EFF}
(S,\sigma,T)/\Lambda_H^4$  at $T/T_C=0.945$  for $\lambda_S=0.001,\,\lambda_{HS}=0,
\,y=0.00172$.
The true minimum is located in the upper right corner.
A shallow local minimum can be seen on the left side.
The red curve is the bounce solution with $S_3/T \simeq141$.}
}
\label{newMIM}
\end{center}
\end{figure}

There is a limitation of our iterative method.
As we have discussed in Section \ref{hQCD-PT-A}, the chiral PT turns into
a cross-over type for large $y$. 
We have found  that our  iterative process
does not converge for large $y$ even much before
the chiral PT turns into a cross-over type.
The reason is that a new  local minimum,
other than the true and false minima, 
develops near the origin $\sigma=S=0$.
The bounce solution passes near the  new  local minimum
to arrive at the origin, as one can see in figure \ref{newMIM}.
The depth of the  
new  local minimum becomes deeper 
and deeper with an increasing value of $y$,
and around a certain value of  $y$ the
 new  local minimum starts to affect the iterative method in such 
 a way that the  iterative process  does not converge.
 At the moment we are not able to find a bounce solution
 beyond this value of $y$.

\section{Gravitational wave spectrum}

There are three production mechanisms of  the stochastic GW
background at a strong first-order PT:
Bubbles are nucleated and
 grow, and then the collisions of the bubble walls  take place,
producing a GW background 
\cite{Kosowsky:1991ua,Kosowsky:1992rz,Kosowsky:1992vn,Kamionkowski:1993fg,Caprini:2007xq,Huber:2008hg}.
We denote by $\Omega_\varphi$ its contribution
to the total GW spectrum  $\Omega_\text{GW}$.
After the bubble-wall collisions 
sound waves  surrounding the bubble walls \cite{Hindmarsh:2013xza,Hindmarsh:2015qta,Giblin:2013kea,Giblin:2014qia,Hindmarsh:2017gnf} 
and magnetohydrodynamic (MHD) turbulence 
\cite{Kosowsky:2001xp,Caprini:2006jb,Kahniashvili:2008pe,Kahniashvili:2008pf,Kahniashvili:2009mf,Caprini:2009yp,Kisslinger:2015hua}
in the plasma become the source of the GW background.
Their contributions to $\Omega_\text{GW}$
are denoted by $\Omega_\text{sw}$
and $\Omega_\text{turb}$, respectively:
\al{
\Omega_\text{GW} (f) h ^2 = 
\left[\,\Omega_\varphi(f) +
\Omega_\text{sw} (f) +\Omega_\text{turb}(f)\,\right] h ^2,
\label{GW-Omega}
}
where $h$ is the dimensionless Hubble parameter,
and $f$ is the frequency of the GW at present.
 To calculate  $\Omega_\text{GW}$ for a given model we 
 first have to find out the nucleation temperature 
$T_n$. Then  we compute the duration time 
of the first-order PT  at  $T=T_n$ and
the released vacuum energy density 
at $T=T_n$.
The released vacuum energy  is indeed the source for the  GW 
energy density, but
 only its part is effectively used as the source.
The corresponding efficiency  is expressed by
 the  efficiency  coefficients that again depend on 
 the released vacuum energy density.
In the following  we start by computing $T=T_n$.

\subsection{ Nucleation temperature $T_n$}

The cosmological tunneling process is quantum mechanical 
transition  from  a false vacuum sate to the true vacuum state
 in the expanding Universe and has
been studied in refs. \cite{Coleman:1977py,Callan:1977pt,Linde:1981zj,Hogan:1984hx,Witten:1984rs}.
The probability of the decay rate of the false vacuum per unit volume per unit time at a finite temperature $T$ is given by \cite{Linde:1981zj}
 \begin{align}
 \label{Gamma}
	\Gamma(T) \simeq T^4 
	\left( \frac{S_3}{2\pi T} \right)^{3/2} \exp(-S_3/T) \,,
 \end{align}
where $S_3$ is the three dimensional Euclidean action 
and is given in eq. (\ref{actionS}) for our model.
The first-order PT proceeds via  the tunneling process
in the expanding Universe, in which
the bubbles of the true vacuum are nucleated.
Since after each tunneling process we have one bubble nucleation,
$\Gamma(T)$ is also  the nucleation rate of the bubbles.
The nucleation temperature $T_n$ is defined 
as the temperature, at which
one bubble for Hubble time and Hubble volume 
is nucleated, i.e., $\Gamma(T_n) / H(T_n)^4=1$, which 
leads to the approximate expression
\cite{Hogan:1984hx,Witten:1984rs}
\be
	\frac{S_3(T_n)}{T_n}& \simeq & 2 \ln \left( \frac{90}{g_*\pi^2}\frac{M^2_\text{Pl}}{T_n^2} \right) \,,
	\label{eq:Nuclcond}
\ee
where we  have ignored the slowly varying factor $(S_3/(2\pi T))^{3/2}$
on the rhs of  eq. (\ref{Gamma}), $g_*$ is the relativistic degrees of freedom
in the Universe at $T=T_n$,
and $M_\text{Pl}=2.435\times 10^{18}$ GeV  is the reduced Planck mass. Then the nucleated bubbles expand
and collide.
Note that the absolute scale of $T_n$ (and also $T_C$) depends crucially
on $\lambda_{HS}$ and $y$, because in the absence of both
couplings the  scale in the hidden sector can be anything;
no information about the energy scale in the SM sector, e.g. $m_h\simeq 0.125$ TeV, can be transferred to the hidden sector.

\subsection{ Duration of the phase transition}
\begin{figure}[t]
\begin{center}
\includegraphics[width=3.4in]{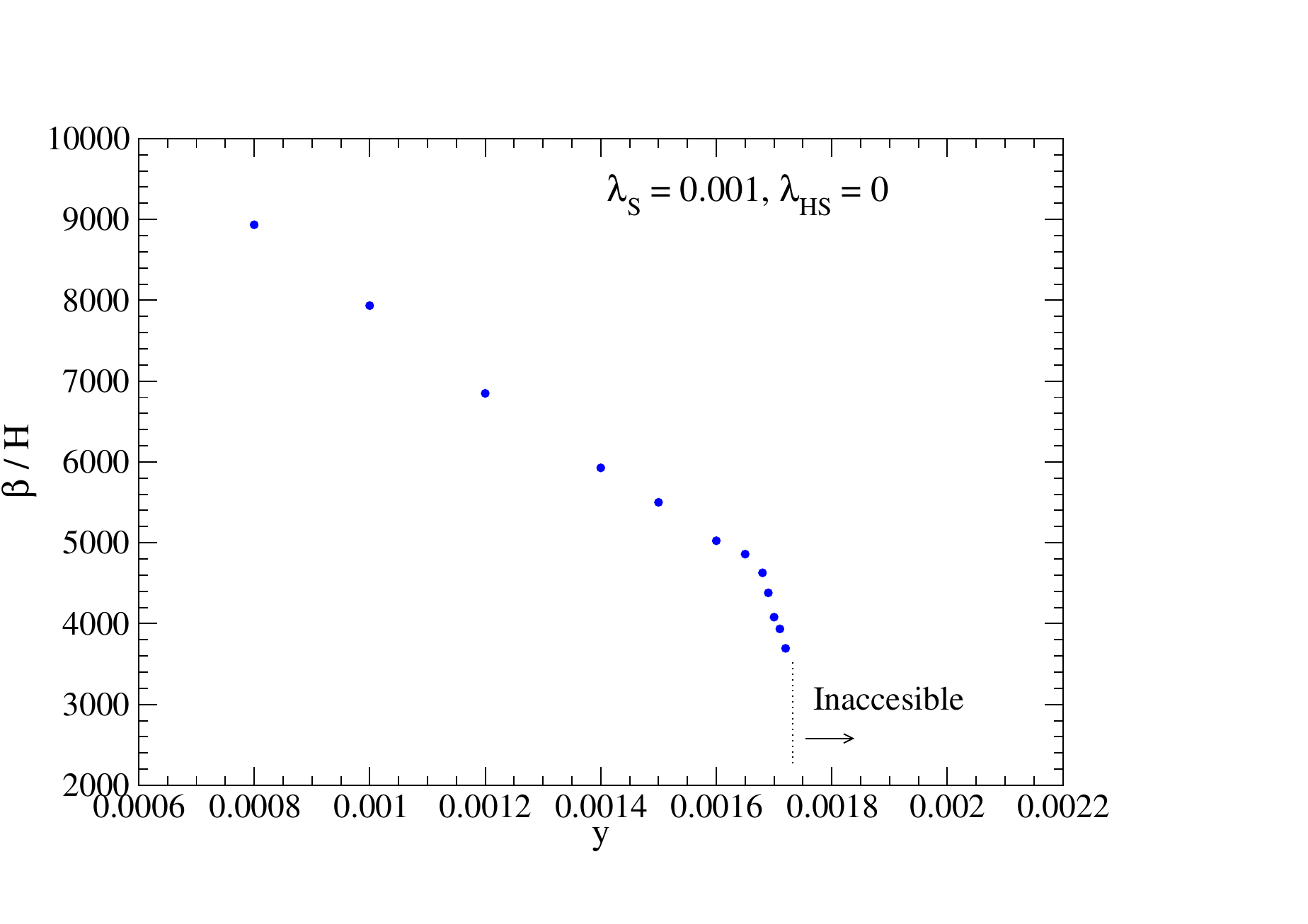}
\hspace{-12mm}
\includegraphics[width=3.4in]{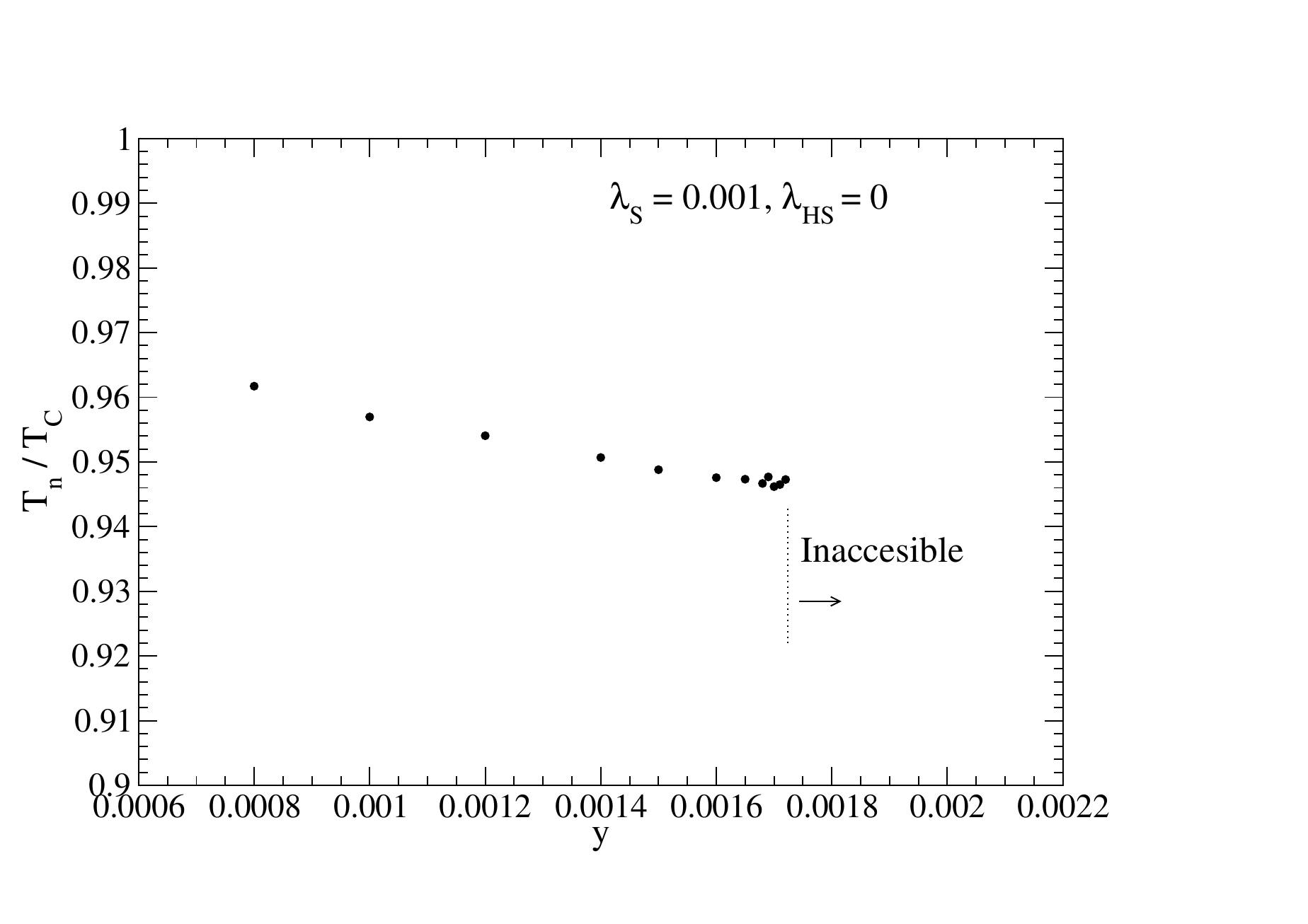}
\vspace{-5mm}
\caption{\footnotesize{Left: $\beta/H$  against the Yukawa coupling $y$ 
for $\lambda_S=0.001$ and $\lambda_{HS}=0$,
which should be compared with $\beta^\text{NJL}/H=
1.4\times 10^4$ \cite{Helmboldt:2019pan} (the value without the singlet scalar $S$).
In contrast to $\beta/H$, $\alpha$ differs only slightly  from the pure
NJL value: $g_*\alpha \simeq 3.8$, where it is  about $3.2$
in the pure NJL case \cite{Helmboldt:2019pan}.
Beyond $y \gtrsim 0.00172$ (for $\lambda_S=0.001$),
the local minimum on the left side in figure \ref{newMIM}
becomes deeper in such a way that the iterative process
in solving the coupled differential equations 
given in (\ref{dfeqs}) and (\ref{dfeqS}) do not converge,
and consequently our method can not be applied.
This area in the parameter space is indicated by ``inaccessible''.
Right: The $y$-dependence of $T_n/T_C$ for $\lambda_S=0.001$ and $\lambda_{HS}=0$.}
}
\label{y-beta}
\end{center}
\end{figure}

The temperature $T$ and time $t$ in the expanding Universe
is related through
\be
\frac{d T}{d t} &=&-H(t) T\,.
\label{t-T}
\ee
The nucleation time $t_n$ is the time, at which 
the temperature $T$ is equal to the nucleation temperature $T_n$.
Since the nucleation time $t_n$  is now defined, 
we can compute the duration of the phase
transition.  To this end, we consider the four-dimensional
Euclidean action 
$S_\text{E}(t)=S_3(T)/T$ and expand it around $t_n$:
\al{
S_E(t)=S_E(t_n)-\beta \Delta t+O((\Delta t)^2)\,,
}
where $\Delta t=(t-t_n)>0$.
Then the nucleation rate for $t\sim t_n$
can be written as
\al{
\Gamma(T)\simeq \Gamma(T_n) e^{\beta \Delta t}.
}
Clearly, the larger $1/\beta$ is, the longer is the time
for which $\Gamma(T)$ stays close to $\Gamma(T_n)$.
Therefore,  $\beta^{-1}$  is  the duration time 
and
can be computed from 
\cite{Kosowsky:1991ua,Kosowsky:1992rz,Kosowsky:1992vn} 
\al{
\beta &= -\frac{dS_E}{dt}\bigg|_{t=t_n} =\frac{1}{\Gamma}
\frac{d \Gamma}{dt}\bigg|_{t=t_n}
= H(t_n) T_n\frac{d}{dT} \left( \frac{S_3}{T} \right)\bigg|_{T=T_n}\,,
\label{beta1}
}
where eqs. (\ref{Gamma}) and (\ref{t-T}) are used.
This means that we need to compute the derivative
of $S_3/T$, which is a cumbersome task in the presence of
many scalar fields involved in the bounce equation like in our case.
To overcome this problem we use the fact that  $S_3/T$
can be well approximated by the fitting function defined in eq. (\ref{fit}).
Since $b$ and $\gamma$ are independent of $x=T/T_C$,
we determine them from the actual calculation of
$S_3/T$ for some $x$ and obtain $\beta/H$ from
\be
\beta/H &=&
 T\frac{d}{dT} 
\,  b (1-T/T_C)^{-\gamma}\bigg|_{T=T_n}
 =b\gamma x_n (1-x_n)^{-1-\gamma}\,,~\mbox{where}~x_n=T_n/T_C\,.
 \label{beta}
\ee
The quantities, $b$ and $\gamma$, do not depend
very much on $\lambda_{HS}$,
 because not only they are dimensionless,
 but also $\lambda_{HS}$ enters into the chiral
 PT  only through the thermal mass of $S$ as one can see in eq. (\ref{thermalM}).
 In contrast to this, they
 and hence $\beta/H$  depend  considerably on $y$,
 because it is the origin of the explicit breaking of the chiral symmetry.
 In figure \ref{y-beta} (left) we show $\beta/H$ for several values of $y$
  with $\lambda_S$ and $\lambda_{HS}$ fixed at 
 $0.001$ and $0$, respectively.
Since $\beta/H\simeq 1.4\times 10^{4}$ in the pure NJL model
\cite{Helmboldt:2019pan},
we see from figure \ref{y-beta} that the larger $y$ is, the more deviation from
the pure NJL model we can expect.
The right panel shows
 the $y$-dependence of $T_n/T_C$ for $\lambda_S=0.001$
  and $\lambda_{HS}=0$, from which we see that in contrast to
  $\beta/H$ the value of $T_n/T_C$ does not change very much
  as $y$ changes.
\footnote{
$\beta/H$ computed in ref. \cite{Aoki:2017aws}
does  not seem to approach the pure NJL value, $\sim 10^{4}$, as the Yukawa coupling
$y$ goes to zero (see, e.g,  the result for the case C in TABLE I and II;
$y=1.07\times 10^{-4}$ but $\beta/H=7.15\times 10^2$.)
Therefore, we suspect that
the modified path deformation method of  ref. \cite{Aoki:2017aws}
to obtain the bounce solution of a coupled system
fails to yield trustful results.}
For  $y\gsim 0.00172$ our iterative method breaks down
(as it is explained in Section \ref{Bounce }), so that
 we stop at $y = 0.00172$ for this example.

\subsection{Released vacuum energy}
As we  see from figure \ref{y-beta}, $\beta/H$ is large
$\sim 10^3$.
Therefore, the scalar contribution  $\Omega_\varphi $ 
to the GW spectrum, being proportional 
to  $(\beta/H)^{-2}$, is much more suppressed than
the other contributions $\Omega_\text{sw} $ 
and $\Omega_\text{turb} $, because they are 
proportional to  $(\beta/H)^{-1}$ (see eqs. (\ref{omephi}),
(\ref{omesw}) and (\ref{ometurb})).
Furthermore, as we will see, the turbulence contribution
$\Omega_\text{turb}$ is suppressed, compared with $\Omega_\text{sw}$,
because the relevant GW frequency $f$ is much
larger than $h_n$, the Hubble parameter at $T_n$, which is  red-shifted today.
Therefore, we here focus on the sound-wave contribution and follow the
treatment of ref. \cite{Espinosa:2010hh}.
It should be noted that the definition of $\alpha$ in 
ref. \cite{Espinosa:2010hh} is not the ratio of the latent heat released at the PT
to the radiation energy of the Universe. Instead, they use the trace of the energy momentum tensor of the plasma, leading to
\be
\alpha &=&\frac{1}{\rho_\text{rad}(T_n)}\left(
\Delta V (T_n)-\frac{1}{4}T\frac{\partial \Delta V(T)}{ \partial T} \bigg|_{T=T_n}\right)\,,
\label{alpha}
\ee
where 
$\Delta V(T)=V_\text{EFF}(0,0,T)-
V_\text{EFF}(\langle S\rangle,\langle\sigma\rangle,T)$, 
and $\rho_\text{rad}(T)=\pi^2 g_* T^4/30$.
According to ref. \cite{Espinosa:2010hh},   if the speed of the wall $\xi_w$ is larger than
$\xi_J$, we may identify the vacuum energy density,
which enters into the definition of $\alpha$, with the vacuum
energy density outside of the bubble (as we have already
done so above),
where $\xi_J $ is the wall speed for the Jouguet detonation
\be
\xi_J &=& \frac{\sqrt{\alpha(2+3\alpha)}+1}{\sqrt{3}(1+\alpha)}\,.
\label{jouguet}
\ee
Correspondingly, $g_*$ in $\rho_\text{rad}$
is the relativistic degrees of freedom
in the symmetry phase, which is 
not necessarily the same as that at the GW production.
So, in our case
\be
g_* =106.75+1+ 8\times 2+\frac{7}{8}\times 3\times 3 \times 2\times 2
=155.25\,.
\label{gstar}
\ee

When calculating
the GW spectrum later on, we will be considering
an optimistic  parameter space given by
\be
\lambda_S &=& 0.001\,, ~\lambda_{HS}
\in [0.0001,0.018]\,, ~y \in [0.0008, 0.00172]\,.
\label{parameter}
\ee
In this parameter space, 
$\alpha$ does not change very much:
\be
0.0242 \lsim\alpha \lsim 0.0250\,.
\label{alpha1}
\ee

\subsection{Reduction of the sound-wave contribution}\label{Reduction}

As we mentioned above, the sound-wave contribution
$\Omega_\text{sw}$ will be the most dominant one in our model.
The formula for $\Omega_\text{sw} h^2$  (see eq. (\ref{omesw}))
has been derived from the 
 numerical simulations for which 
a long-lasting  source of the GW,
 i.e., $\tau_\text{sw}H > 1$,  is assumed  \cite{Hindmarsh:2017gnf},
where 
\be
\tau_\text{sw}\simeq (8\pi)^{1/3}  \frac{\xi_w}{\bar{U}_f \beta}
\label{tausw} 
\ee
 is  the duration of the sound-wave period, $\bar{U}_f$ is the root-mean four-velocity of the plasma, and $\xi_w$ stands for the speed of the wall
 ($\beta$ is defined in eq. (\ref{beta})).
That is, $\tau_\text{sw} H\propto (\beta/H)^{-1}$,
so that $\tau_\text{sw} H > 1$ is unlikely satisfied in our model, because
$\beta/H \gtrsim O(10^3)$.
In refs. \cite{Ellis:2018mja,Ellis:2019oqb} it has been suggested,
for the case that $\tau_\text{sw} H < 1$, to 
use this quantity  as a reduction factor for 
$\Omega_\text{sw} $ to take into account the fact that
the sound wave is an active GW source only for a period shorter
 than the Hubble time. Here we follow ref. \cite{Ellis:2018mja} 
 along with  ref. \cite{Espinosa:2010hh} to calculate 
 $\tau_\text{sw}$ and
 consider  throughout the case of detonations of the 
 plasma motion.

\begin{figure}[t]
\begin{center}
\includegraphics[width=3.7in]{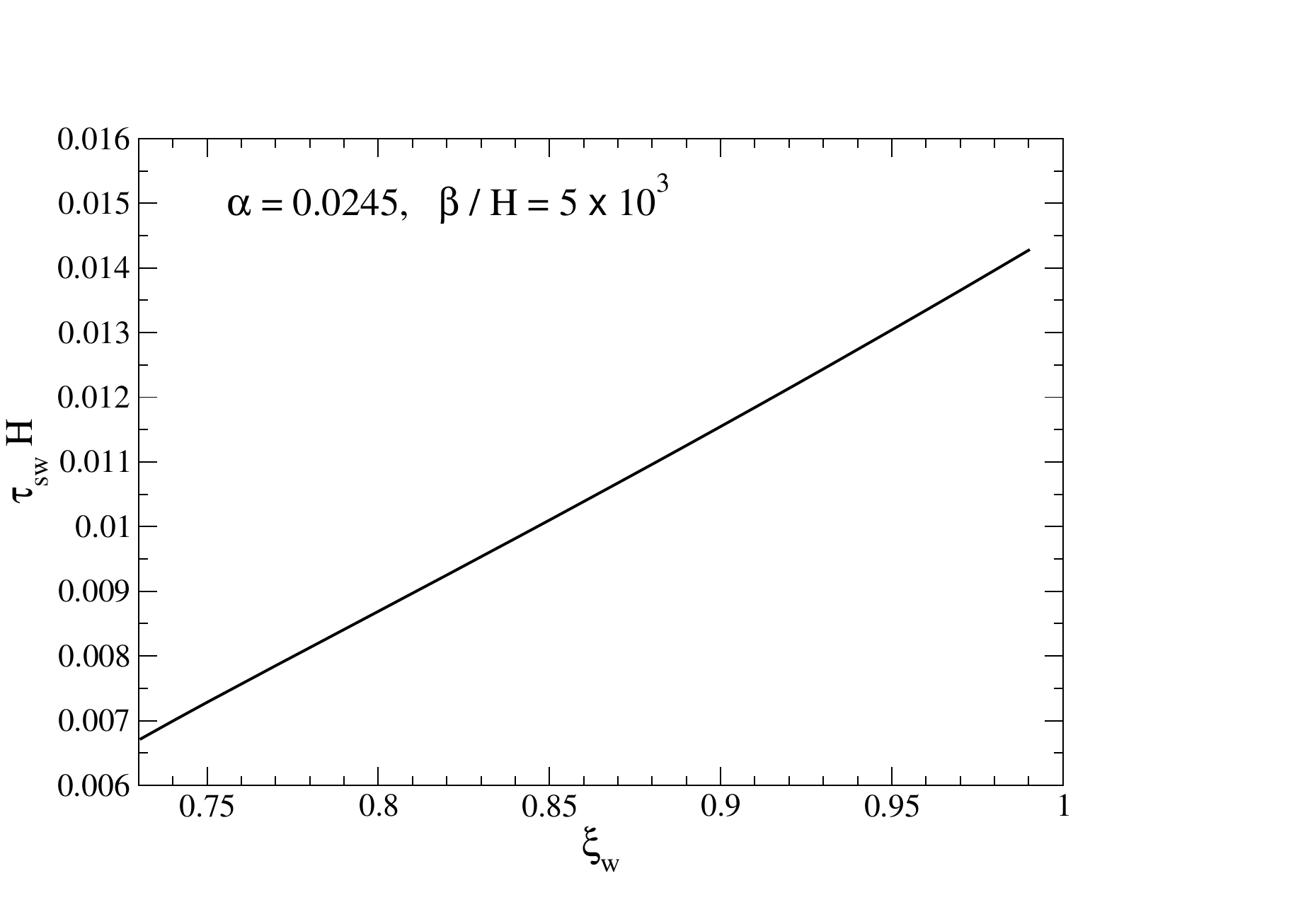}
\vspace{-5mm}
\caption{\footnotesize{The reduction factor $\tau_\text{sw} H$ for $\Omega_\text{sw}$
against the wall speed 
$\xi_w \geq \xi_J\simeq 0.691$ for  $\alpha=0.0245$ and $\beta/H=5\times 10^3$,
where $\tau_\text{sw}$ is defined in (\ref{tausw}).}
}
\label{fig:tausw}
\end{center}
\end{figure}

 The root-mean four-velocity $\bar{U}_f$ can be calculated from 
 \cite{Espinosa:2010hh}
 \be
\bar{U}^2_f
&=&\frac{3}{\xi_w^3}\int_{c_s}^{\xi_w} d\xi
\frac{\xi^2 v^2(\xi)}{1-v^2(\xi)}\,,
\ee
where $v(\xi)$ is the velocity profile of the plasma
in the frame of the bubble center, and $c_s$ is the speed of sound
in the plasma (we assume here $c_s=1/\sqrt{3}$).
The velocity profile $v(\xi)$ satisfies the first-order differential equation
 \cite{Espinosa:2010hh}
\be
\frac{v}{\xi}
&=&\frac{1}{2}\left(\frac{1-v\,\xi}{1-v^2}\right)\left(\frac{\mu^2(\xi,v)}{c_s^2}-1\right)
\frac{d v}{d\xi}\,,~\mbox{where}~
\mu(\xi,v)=\frac{\xi-v}{1-\xi v}\,.
\label{profile}
\ee
To solve the differential equation  (\ref{profile}) uniquely, we use $v(\xi_w)$ 
as an initial value, i.e., the plasma speed
just behind the wall. 
Since we focus on the detonations, 
the plasma in front of the wall is at rest in the bubble center frame, i.e.,
$v_+=\xi_w$, where $v_+$ is the speed of the plasma
in front of the wall in the wall frame ($\xi_w$ is the speed of the wall
 in the bubble center frame).
Therefore, the speed of the plasma just behind the wall
in the wall frame, denoted by $v_-$, can be obtained by  the Lorentz transformation
\be 
-v_- &=& \frac{v(\xi_w)-\xi_w}{1-\xi_w v(\xi_w)}\,.
\label{vm}
\ee
(The minus sign is introduced, because 
the plasma velocity in the wall frame has  the opposite
direction compared with the wall velocity in the 
bubble center frame.)
Eq. (\ref{vm}) can be used to obtain
\be
v(\xi_w) &=& \frac{v_+ -v_-}{1-v_+ v_-}
~~\mbox{with} ~v_+=\xi_w\,,
\label{vxiw}
\ee
where $v_\pm$ are constrained by the matching equations
between the plasma sates in front of and behind the wall:
\be
\xi_w &=&v_+=
\frac{1}{1+\alpha}\left[
\left(\frac{v_-}{2} +\frac{1}{6 v_-}\right) + \left\{\left(\frac{v_-}{2} +\frac{1}{6 v_-}\right) ^2 +
\alpha^2+ \frac{2}{3}\alpha -\frac{1}{3}\right\}^{1/2}
\right]\,.
\label{xiw}
\ee
So, we obtain  $v_-$ from eq. (\ref{xiw})  for a given set of
$\xi_w$ and $\alpha$ and insert it into the rhs of eq. (\ref{vxiw}) to
obtain the initial value $v(\xi_w)$.
Since the minimum value of $v_-$ is $c_s$ for the detonations
to be realized  \cite{Espinosa:2010hh}, we find that the minimum value of
$\xi_w$ is just the Jouguet speed $\xi_J$
defined in eq. (\ref{jouguet}).
To obtain an idea on the size of $\tau_\text{sw} H$, we show 
$\tau_\text{sw} H$ in figure \ref{fig:tausw}  as a function of $\xi_w ~(\geq \xi_J)$
for $\alpha=0.0245$ and $\beta/H=5\times 10^3$.
As we see from figure \ref{fig:tausw}, 
the reduction factor
for $\Omega_\text{sw}$ is  of order $10^{-2}$ in our model.

\subsection{Gravitational wave spectrum}
Now we are in position to present the GW spectrum
$\Omega_{\mathrm{GW}}$ of our model.
As we have argued, the area (\ref{parameter}) is an optimistic choice
of the parameter space,
and we expect that $\Omega_{\mathrm{GW}}$ 
will be  smaller in other regions of the parameter space.
The relativistic  degrees of freedom in the expanding Universe
enters in the following expressions.
It is the  relativistic  degrees of freedom $g'_*$ at the time,
at which the GW background is produced.
Therefore, $g'_*$  varies with the time,
because the tunneling process takes place for a finite period of time.
It is certainly not $g_*$ that is the one in the symmetric phase (\ref{gstar})
and has been used for the computation of $\alpha$ in eq. (\ref{alpha}).
In the following we assume that $g'_*$  can be  approximated
by the relativistic  degrees of freedom in the broken phase:
\be
g'_* &=&106.75+8+1+1\,,
\ee
where $8$ comes from the NG bosons, and $1$ is from $\sigma$ as well as from $S$.

Numerical simulations and  analytic estimates
 \cite{Kosowsky:1991ua,Kosowsky:1992rz,Kosowsky:1992vn,Kamionkowski:1993fg,Caprini:2007xq,Huber:2008hg,Jinno:2019bxw,Hindmarsh:2013xza,Hindmarsh:2015qta,Giblin:2013kea,Giblin:2014qia,Hindmarsh:2017gnf,Kosowsky:2001xp,Caprini:2006jb,Kahniashvili:2008pe,Kahniashvili:2008pf,Kahniashvili:2009mf,Caprini:2009yp,Kisslinger:2015hua}
 of the individual contributions
to $\Omega_\text{GW}$
 lead to the following formula:
\begin{itemize}
  \item Scalar field contribution $\Omega_{\varphi}$ \cite{Huber:2008hg}:
 \begin{align}
h^2\,\Omega_{\varphi}(f)=1.67\times 10^{-5}
 (\beta/H)^{-2}\left(\frac{\kappa_{\varphi} \alpha}{1+\alpha}\right)^2
\left(\frac{100}{g'_{*}}\right)^{1/3}\left(\frac{0.11\xi^3_{w}}{0.42+\xi^2_w }\right)S_{\varphi}(f),
 \label{omephi}
\end{align}
 where 
  \begin{align}
  \label{Sphi}
S_{\varphi}(f) & =\frac{3.8(f/f_{\varphi})^{2.8}}{1+2.8(f/f_{\varphi})^{3.8}}
\end{align}
with the peak frequency 
 \begin{align}
 \label{fpeakphi}
  f_{\varphi} & = 16.5 \times 10^{-6} (\beta/H) \left(\frac{0.62}{1.8-0.1\xi_w+\xi^2_w }\right)\left( \frac{T_n}{100~\mathrm{GeV}}\right)\left( \frac{g'_{*}}{100}\right)^{1/6}~\mathrm{Hz}.
\end{align}
  \item Sound-wave contribution $\Omega_{\mathrm{sw}}$
 \cite{Hindmarsh:2013xza,Hindmarsh:2015qta}:
 \begin{align}
h^2\,\Omega_{\mathrm{sw}}(f)=(\tau_\text{sw} H)\,2.65\times 10^{-6} 
\,(\beta/H)^{-1}\left(\frac{\kappa_\text{sw} \alpha}{1+\alpha}\right)^2\left(\frac{100}{g'_{*}}\right)^{1/3}\xi_{w}S_{\mathrm{sw}}(f),
 \label{omesw}
\end{align}
 where 
 \begin{align}
  \label{Ssw}
S_{\mathrm{sw}}(f) & =(f/f_{\mathrm{sw}})^3\left(\frac{7}{4+3(f/f_{\mathrm{sw}})^2}\right)^{7/2}
\end{align}
with the peak frequency
 \begin{align}
 \label{fpeak}
  f_{\mathrm{sw}} & = 1.9 \times 10^{-5} \xi_w ^{-1} 
  (\beta/H)\left( \frac{T_n}{100~\mathrm{GeV}}\right)\left( \frac{g'_{*}}{100}\right)^{1/6}~\mathrm{Hz}.
\end{align}
According to refs. \cite{Ellis:2018mja,Ellis:2019oqb},  the reduction factor  
$\tau_\text{sw} H$
(calculated  in Section \ref{Reduction}) is multiplied in eq. (\ref{omesw}).
  \item MHD turbulence contribution $\Omega_{\mathrm{turb}}$
  \cite{Caprini:2009yp}:
 \begin{align}
h^2\,\Omega_{\mathrm{turb}}(f)=(1-\tau_\text{sw}H)\,3.35\times 10^{-4} 
\,(\beta/H)^{-1}\left(\frac{\kappa_\text{sw}\alpha}{1+\alpha}\right)^{\frac{3}{2}}\left(\frac{100}{g'_{*}}\right)^{1/3}
\xi_{w}S_{\mathrm{turb}}(f),
 \label{ometurb}
\end{align}
 where 
 \begin{align}
  \label{Sturb}
S_{\mathrm{turb}}(f) & =\frac{(f/f_{\mathrm{turb}})^3}{[
1+(f/f_{\mathrm{turb}})]^{\frac{11}{3}}(1+8\pi f/h_n)}
\end{align}
with the peak frequency
 \begin{align}
 \label{fpeakturb}
  f_{\mathrm{turb}} & = 2.7 \times 10^{-5} \xi_w ^{-1} 
  \tilde{\beta}\left( \frac{T_n}{100~\mathrm{GeV}}\right)\left( \frac{g'_{*}}{100}\right)^{1/6}~\mathrm{Hz}\,,
\end{align}
 and  
\begin{align}
  \label{h}
h_n & = 16.5 \times 10^{-6} \left( \frac{T_n}{100~\mathrm{GeV}}\right)\left( \frac{g'_{*}}{100}\right)^{1/6}~\mathrm{Hz}\,,
\end{align}
which is the value (redshifted to today)  of the  Hubble parameter
 at the production of the GW. We have introduced the   enhancement factor
$(1-\tau_\text{sw}H)$ in eq. (\ref{ometurb}) and used the same efficiency coefficient
as for the sound-wave contribution \cite{Ellis:2019oqb}.
\end{itemize}

As we have mentioned in various places and we can see now from eq. (\ref{omephi}),
the scalar contribution $\Omega_\varphi$ is,
 due to $\beta/H\sim 10^3$, about 2 orders of magnitude
smaller than $\Omega_\text{sw}$.
Furthermore, the case at hand corresponds to a nonrunaway scenario,
in which the friction between the bubbles in the surrounding plasma 
prevents the acceleration of the  bubble  expansion
\cite{Bodeker:2009qy,Espinosa:2010hh}.
To see this, we estimate $\alpha_\infty$ according to  ref. \cite{Espinosa:2010hh}:
\be
\alpha_\infty
&\simeq &
\frac{30}{24 \pi^2 g_* T_n^2}\left[\,
\frac{1}{2}n_f n_c M^2(\langle S \rangle,\langle \sigma \rangle)+
M_S^2(\langle S \rangle)
\,\right] \in (0.078,0.098)
\ee
for the parameter space (\ref{parameter}),
where $M(S , \sigma)$ and $M_S(S)$ are given in 
eqs.  (\ref{M}) and (\ref{thermalM}),
respectively.
Therefore, $\alpha_\infty < \alpha\simeq 0.024$
(see (\ref{alpha1})), so that we have a 
 nonrunaway scenario \cite{Espinosa:2010hh} and  ignore the scalar contribution
 (\ref{omephi}) in the following discussion.

We use the efficient coefficient $\kappa_\text{sw}$ given in ref. \cite{Espinosa:2010hh}
for $\Omega_\text{sw}$ and also for $\Omega_\text{turb}$:
\be
\kappa_\text{sw}(\xi_w \gsim \xi_J)
&\simeq&
\frac{\chi_J^3\,(\xi_J/\xi_w)^{5/2}\,\kappa_C \kappa_D}
{( \chi_J^3-\chi_w^3)\,\xi_J^{5/2} \kappa_C+\chi_w^3\kappa_D}\,,
\ee
where
\be
\chi_J &=& \xi_J-1 \, ,\,\chi_w=\xi_w-1\,,\nn\\
\kappa_C &\simeq&  \frac{\alpha^{1/2}}{0.135+0.98^{1/2}+\alpha}\,,\,
\kappa_D \simeq \frac{\alpha}{0.73+0.083 \,\alpha^{1/2}+\alpha}\,,
\ee
and $\xi_J$ is given in eq. (\ref{jouguet}).
Although $\Omega_\text{sw}$ is reduced by the reduction factor
$\tau_\text{sw} H$ and $\Omega_\text{turb}$ is enhanced
by $(1-\tau_\text{sw} H)$ and also by the identification 
$\kappa_\text{turb}=
\kappa_\text{sw}$,
the turbulence contribution  $\Omega_\text{turb}$ is about one order of magnitude smaller than 
$\Omega_\text{sw}$, because $f_\text{turb}/h_n\sim \beta/H\sim 10^3$
 that is in the denominator of eq. (\ref{Sturb});
 $(f_\text{turb}/h_n)^{-1}/(\tau_\text{sw} H) \sim 0.1$.

\begin{figure}[t]
\begin{center}
\includegraphics[width=3.3in]{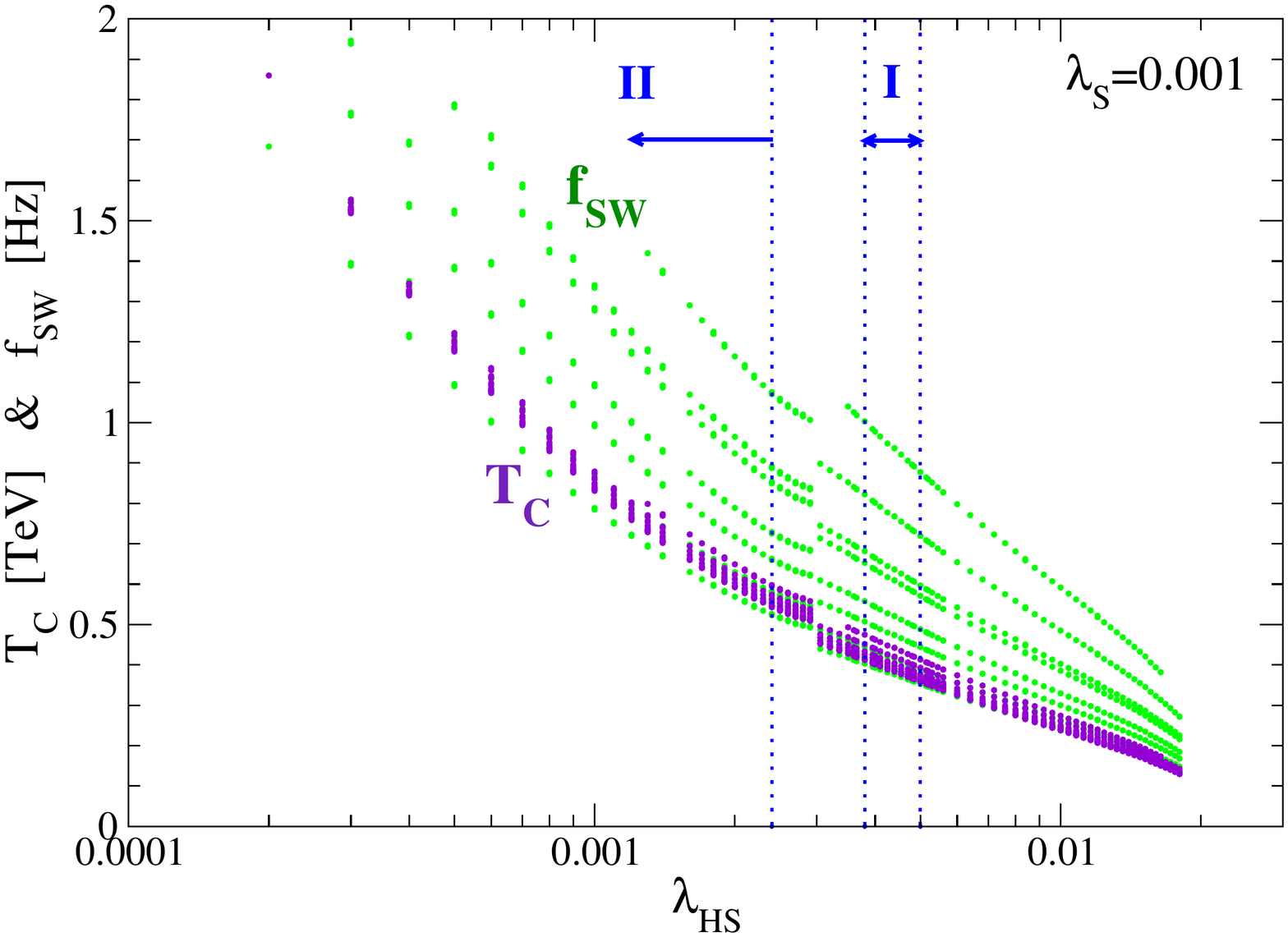}
\hspace{-6mm}
\includegraphics[width=3.3in]{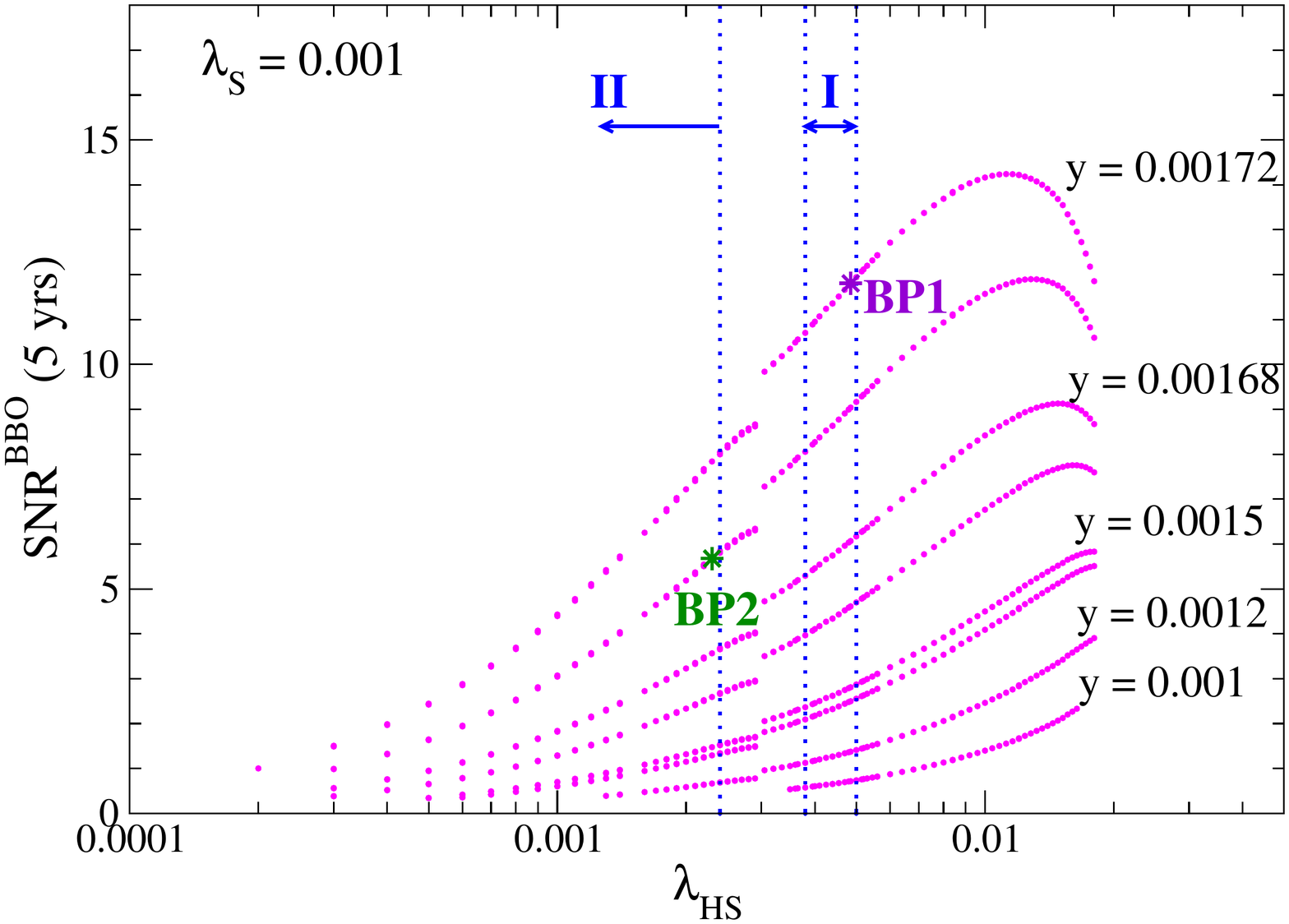}
\vspace{-5mm}
\caption{\footnotesize{
Left: The critical temperature $T_C $ [TeV] (purple)
and the peak frequency of the sound-wave contribution 
$f_\text{sw}$ [Hz]  (green) against  $\lambda_{HS}$.
The area I and II are allowed by LHC (see figure \ref{sint-ms-mDM}).
Right: 
SNR$^\text{BBO}$ against $\lambda_{HS}$  with  five years  observation.
 The SNR$^\text{BBO}\,(5\, \text{yrs})$ of the benchmark points, 
BP1(purple star) and BP2 (green star), 
are also plotted.}}
\label{rhs-Tc}
\end{center}
\end{figure}

As we see from eq. (\ref{fpeak}) the scale of the GW frequency  is fixed by the
nucleation temperature $T_n$. Note that the absolute scale 
of the critical temperature $T_C$ and hence $T_n$ is fixed
through the coupling with the SM sector, i.e., $\lambda_{HS}$ and $y$. 
In the left panel of figure \ref{rhs-Tc} we show $T_C$ [TeV] and 
$f_\text{sw}$ [Hz]
against $\lambda_{HS}$. 
Obviously, the smaller $\lambda_{HS}$ is, the larger is $\Lambda_H$
(the scale of the hidden sector), and consequently
  higher  $T_C$  and $f_\text{sw}$.
  The band of $f_\text{sw}$ is wider than that of $T_C$, because
  $\beta/H$ depends on $y$ (see figure \ref{y-beta}) more than
 $T_C$ does. 
 As we also see from this figure that the GW frequencies 
in our model are  $\gsim 0.3$ Hz, which can be covered by 
DECIGO \cite{Seto:2001qf,Kawamura:2006up,Kawamura:2011zz}
and BBO \cite{BBO,Cornish:2005qw,Corbin:2005ny}.

We calculate the SNR according to ref. 
\cite{Thrane:2013oya},
\al{
	\text{SNR}=\sqrt{2 t_{\text{obs}} \int_{f_{\text{min}}}^{f_{\text{max}}} \!\!  df \, \bigg[
\frac{\Omega_\text{GW}(f)\, h^2}
{\Omega_\text{noise}(f)\, h^2}\bigg]^2},
	\label{eq:SNR}
}
where $t_{\text{obs}}$ stands for
 the duration of an observation in seconds,  and  $(f_\text{min},f_\text{max})$  is the frequency range of a given experiment.
The quantity $\Omega_\text{noise}(f)\,h^2$ represents the effective strain noise power spectral density for a given detector network, expressed as energy density parameter \cite{Moore:2014lga}.
For the space-based observatories mentioned above, we adopt the strain noise power spectral densities from refs. \cite{Yagi:2011yu,Yagi:2013du,Isoyama:2018rjb}.
(We use the sky-averaged sensitivity \cite{Yagi:2013du}.)
The result\footnote{The SNR is computed including the 
 turbulence contribution. }, SNR against $\lambda_{HS}$ for BBO, 
 is shown in the right panel of figure \ref{rhs-Tc}, 
 \footnote{
 The effect of unresolvable astrophysical foregrounds from black hole, 
 neutron star and white dwarf mergers on the signal significance
 are ignored.}
where we assume that $t_{\text{obs}}=5$ years and the speed of the wall 
 $\xi_w$ is equal to the Jouguet speed $\xi_J$ given 
 in eq. (\ref{jouguet}). 
 The SNR$^\text{BBO}\,(5\, \text{yrs})$ of the benchmark points, BP1 and BP2
 defined in (\ref{BP}), are
 $11.8$ and $5.7$,  respectively, while for DECIGO we find
SNR$^\text{DECIGO}\,(5\, \text{yrs})=1.1$ and $0.5$, respectively.
 Therefore, there is a good chance that the GW signals of our model can be detected
 by BBO, where the area I and II are allowed by LHC  (see figure \ref{sint-ms-mDM}).

\begin{figure}[t]
\begin{center}
\includegraphics[width=3.3in]{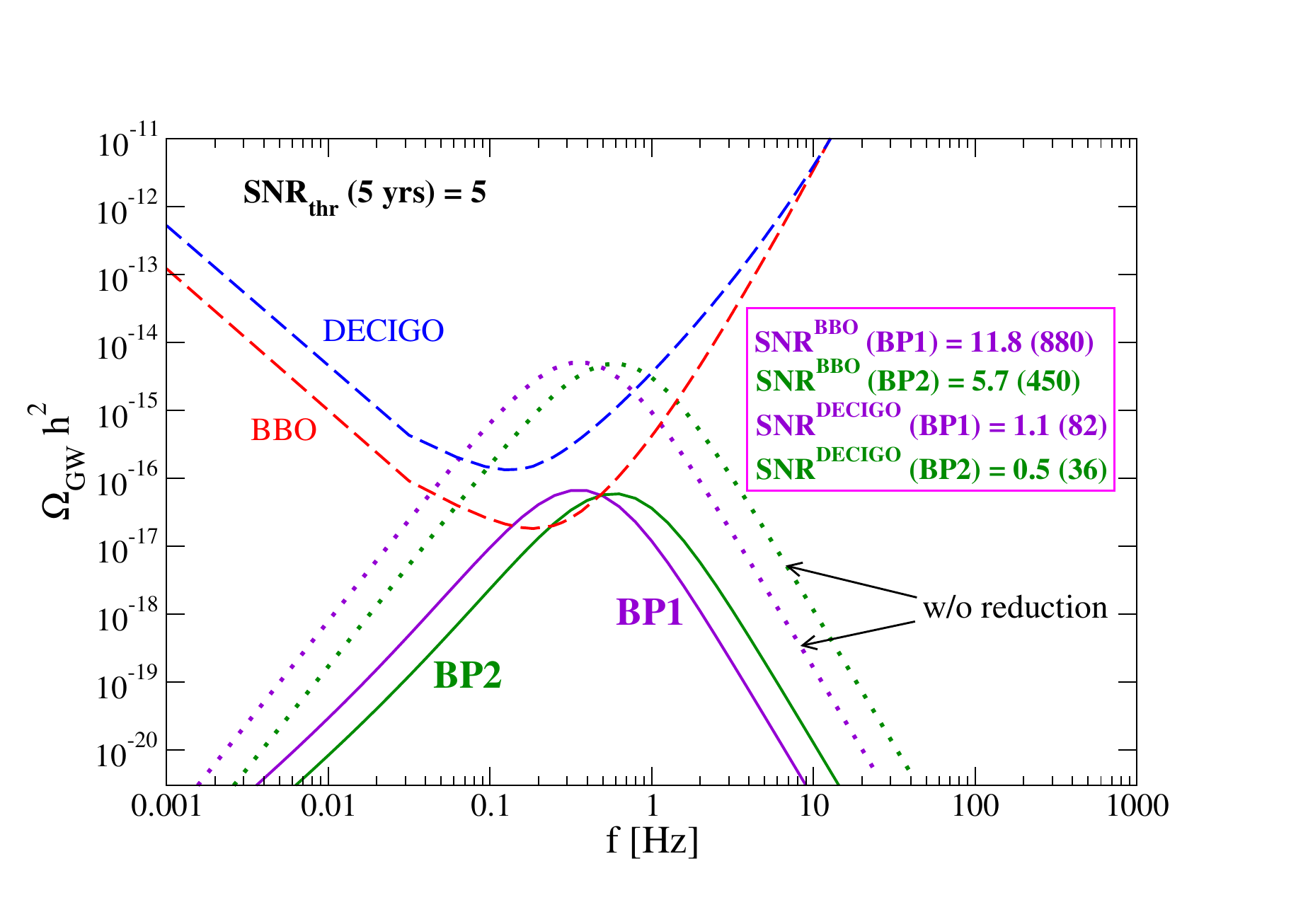}
\hspace{-6mm}
\includegraphics[width=3.3in]{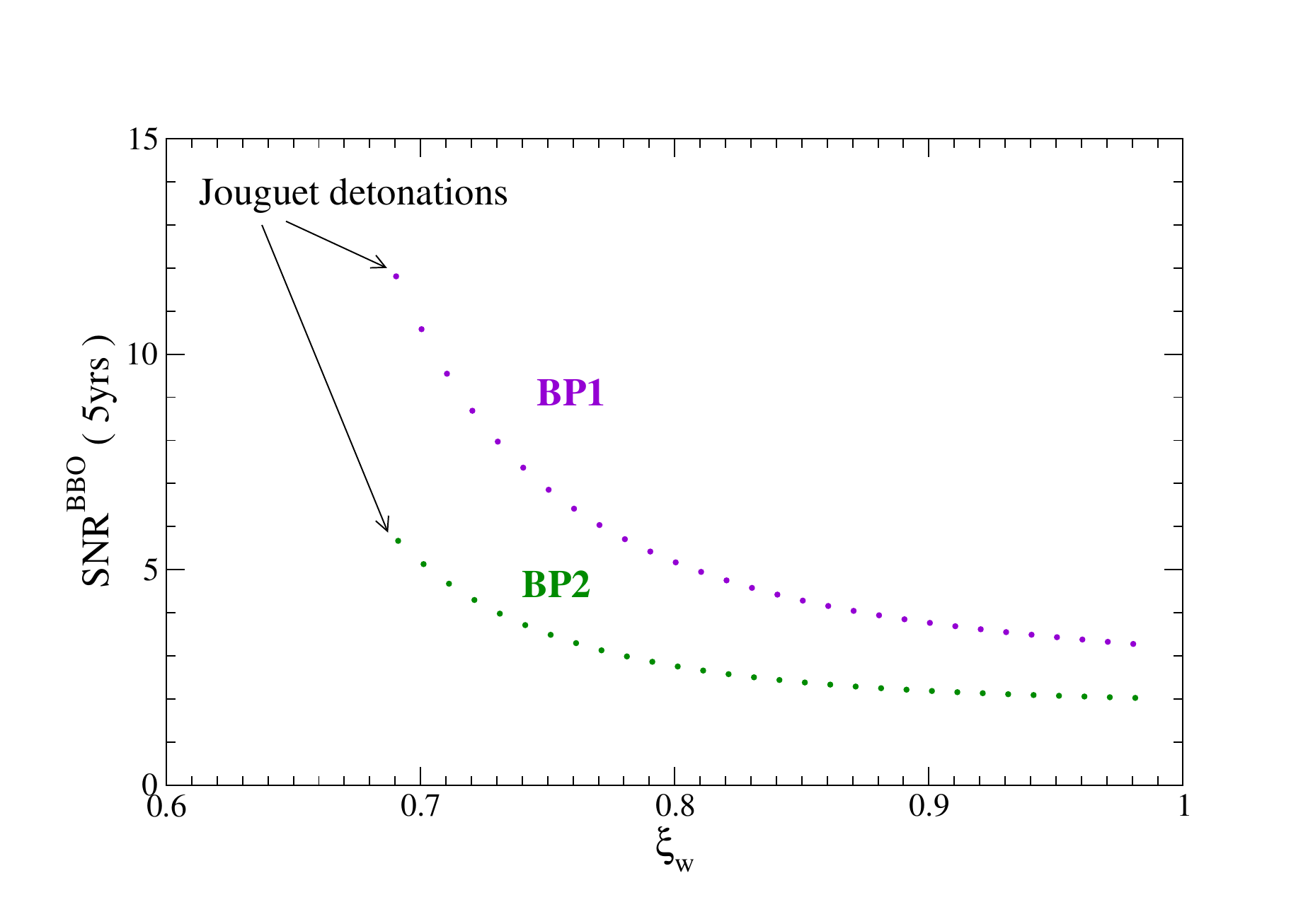}
\vspace{-5mm}
\caption{\footnotesize{Left: The GW spectrum for the benchmark points BP1 (purple),
BP2 (green) and
the power-law-integrated sensitivity of BBO (red dashed curve)
as well as  DECIGO (blue dashed curve), 
 where we assume that the threshold SNR is $5$
($\rho_\text{thr}=5$) with five years 
observation for both detectors. The GW spectrum is computed including the turbulence contribution, which is about one order of magnitude smaller
 than that of the sound-wave contribution. 
 The dotted purple and green lines present, respectively,
  the GW spectrum of BP1 and BP2, for which the reduction factor
$\tau_\text{sw} H$   due to the short sound-wave period is ignored.
Right:  The $\xi_w$ dependence of SNR$^\text{BBO}\,(5\, \text{yrs})$.
The  Jouguet speed $\xi_J$ is the minimum speed of $\xi_w$
for detonations. At this speed the SNR becomes maximal.}}
\label{GWspectrum}
\end{center}
\end{figure}

In the left panel of
figure \ref{GWspectrum} we present the GW spectra for BP1 (purple) and BP2 (green) 
with $\xi_w=\xi_J$, which should be compared with 
the power-law-integrated sensitivity \cite{Thrane:2013oya} of BBO (red dashed curve)
and DECIGO (blue dashed curve), 
 where we assume that the threshold SNR is $5$
($\rho_\text{thr}=5$) with five years observation for both detectors.
Since a part of the spectral curves for BP1 and BP2
runs over the sensitivity curve of BBO,
we see once again that their signals  could be detected
at BBO, while for DECIGO it would be very difficult.
For comparison we also present the GW spectra
 (dotted purple and green lines), which we obtain without  
 the reduction factor $\tau_\text{sw} H$.
We see a difference of 2 orders of magnitude, whose origin 
is  nothing but $\tau_\text{sw} H\sim 10^{-2}$.

As the last task we consider the dependence of the wall speed $\xi_w$,
because we have assumed so far that it is equal to the Jouguet speed
$\xi_J$.
In the right panel of figure \ref{GWspectrum} we show the
$\xi_w$ dependence of SNR$^\text{BBO}\,(5\, \text{yrs})$.
In fact, SNR$^\text{BBO}\,(5\, \text{yrs})$ assumes the maximal value
at $\xi_w=\xi_J$, which follows from the fact that the reduction factor
$\tau_\text{sw} H$ decreases as $\xi_w$ increases
(see figure \ref{fig:tausw}).
But there is still a sufficient range in the parameter space,
in which the detectability threshold is exceeded.

\section{Summary and Conclusion}

In this paper we have studied the stochastic GW background produced 
at the cosmological chiral PT in 
a conformal extension of the SM 
\cite{Kubo:2014ida,Ametani:2015jla} 
and extended the analysis of  ref. \cite{Aoki:2017aws}.
In particular, we have re-calculated $\beta/H$, because
$\beta/H$ in ref. \cite{Aoki:2017aws}
does not approach the pure NJL value, $\sim 10^{4}$, as the Yukawa coupling $y$ decreases and for this reason
we have suspected that
the modified path deformation method of  ref. \cite{Aoki:2017aws}
to obtain the bounce solution of a coupled system
fails to yield trustful results.

Therefore, we have adopted
an iterative method (with a reasonable convergence property) and found
that $S_3/T$ 
 can be fitted with a simple function (\ref{fit}).
Using this fitting function for the determination of $\beta/H$
we have obtained $\beta/H \simeq (4-9)\times 10^3$
in the optimistic parameter space.
We also have found that the benchmark point values of
$\beta/H$ presented in ref. \cite{Aoki:2017aws} are 
about one order of magnitude
smaller than those calculated by using the new method.

There are, in the $SU(3)_V$ flavor symmetry limit, five independent
 parameters, $\lambda_H,
\lambda_S, \lambda_{HS}, y$ and $g_H$ (or the hidden sector scale
$\Lambda_H$), where effectively two of them are used to obtain
$m_h=125$ GeV and $\langle h\rangle=246$ GeV.
We have systematically narrowed the parameter space,
giving smaller values of $\beta/H$ than that of the pure NJL model 
 and hence
 larger (dimensionless) spectral GW energy density  
$\Omega_\text{GW}$.
Obviously, $\Omega_\text{GW}$
 will be smaller in other regions of the parameter space. 
 In this optimistic parameter space (with $\lambda_S\sim 10^{-3}$)  
the singlet scalar $S$ 
can become as light as the Higgs  $h$, and therefore we have taken into account the LHC constraint on their mixing:
There are two allowed regions for $\lambda_S=0.001$ 
 that are denoted by I (for $m_S <  m_h$) and II (for $m_S >  m_h$). 
We remark that for this optimistic parameter space  in the $SU(3)_V$ flavor symmetry limit no realistic DM relic abundance can be obtained
because the  resonant condition  
($ m_S \simeq 2m_{\rm DM}$) in the s-channel of the DM annihilation 
can not be realized \cite{Holthausen:2013ota}.
(In the model studied in ref. \cite{Aoki:2017aws} 
the flavor group $SU(3)_V$ is unbroken and the
hidden fermions have no $U(1)_Y$ charge.)
This is why we have considered the model with 
a finite $U(1)_Y$ charge for the hidden fermions and
have explicitly broken $SU(3)_V$ down to $SU(2)_V\times U(1)$
to apply the mechanism of ref. \cite{Ametani:2015jla} to obtain
a realistic DM relic abundance.
But for the analyses of the GW background spectrum 
we have considered the $SU(3)_V$ limit, because
it is only marginally broken and we would have had to deal with
three variables (instead of two) to find a bounce solution.

The fact, $\beta/H \simeq (4-9)\times 10^3$,
implies a short duration time of the first-order
chiral PT, much shorter than the Hubble time, and consequently a short sound wave period $\tau_\text{sw}$ as  an active GW source;
 $\tau_\text{sw} H \sim10^{-2}$.  
 Then following refs. \cite{Ellis:2018mja,Ellis:2019oqb}
we have used $\tau_\text{sw} H$ as  the reduction factor for the sound wave contribution
$\Omega_\text{sw}$, which  is nevertheless
 the most dominant contribution
to $\Omega_\text{GW}$.
We have evaluated the SNR for DECIGO and BBO and found
that SNR$^\text{DECIGO} \lsim 1.2$ and 
SNR$^\text{BBO} \lsim 12.0$ with five years observation,
from which we conclude that
the GW signal predicted by the model in the optimistic case could be detected at BBO.
\footnote{With a new sensitivity introduced in ref. (\cite{Alanne:2019bsm})
the detectability may increase.}

At last we recall that the results obtained by using
effective theory methods to study the GWs produced 
at a first-order PT in
strong-interacting QCD-like theories
 agree  with each other only qualitatively \cite{Helmboldt:2019pan} and
 for a more precise determination of the GWs
we need  first-principle calculations (like lattice simulations)
 which may become available in future \cite{Brower:2019oor}.

\section*{Acknowledgments}
We thank Alexander J. Helmboldt and Susan van der Woude for useful discussions.
J.~K.~thanks 
for kind hospitality at the Max-Planck-Institute for
Nuclear Physics, Heidelberg,
where a part of this work has been done.
The work of M.~A. is supported in part by the Japan Society for the
Promotion of Sciences Grant-in-Aid for Scientific Research (Grant
No. 17K05412). 
J.~K.~is partially supported by the Grant-in-Aid for Scientific Research (C) from the Japan Society for Promotion of Science (Grant No.19K03844).

\bibliographystyle{JHEP}
\bibliography{gw-2019}

\providecommand{\href}[2]{#2}\begingroup\raggedright\begin{thebibliography}{100}

\bibitem{Lindner2019}
M.~Lindner, {\it {Conformal extensions of the Standard Model}},  in {\em
  {Particle Astrophysics and Cosmology Including Fundamental InteraCtions
  (PACIFIC 2019), Moore, French Polynesia, September 1-6, 2019}}, 2019.
\newblock
  \href{http://www.arxiv.org/abs/https://conferences.pa.ucla.edu/pacific-2019/talks/lindner.pdf}{{\tt
  https://conferences.pa.ucla.edu/pacific-2019/talks/lindner.pdf}}.

\bibitem{Ross:2017oiz}
G.~Ross, {\it {Beyond the Standard Model}},  {\em Phys. Atom. Nucl.} {\bf 79}
  (2016), no.~11-12 1445--1470.

\bibitem{Holthausen:2011aa}
M.~Holthausen, K.~S. Lim, and M.~Lindner, {\it {Planck scale Boundary
  Conditions and the Higgs Mass}},  {\em JHEP} {\bf 02} (2012) 037,
  [\href{http://www.arxiv.org/abs/1112.2415}{{\tt 1112.2415}}].

\bibitem{Bezrukov:2012sa}
F.~Bezrukov, M.~{\relax Yu}. Kalmykov, B.~A. Kniehl, and M.~Shaposhnikov, {\it
  {Higgs Boson Mass and New Physics}},  {\em JHEP} {\bf 10} (2012) 140,
  [\href{http://www.arxiv.org/abs/1205.2893}{{\tt 1205.2893}}]. [,275(2012)].

\bibitem{Degrassi:2012ry}
G.~Degrassi, S.~Di~Vita, J.~Elias-Miro, J.~R. Espinosa, G.~F. Giudice,
  G.~Isidori, and A.~Strumia, {\it {Higgs mass and vacuum stability in the
  Standard Model at NNLO}},  {\em JHEP} {\bf 08} (2012) 098,
  [\href{http://www.arxiv.org/abs/1205.6497}{{\tt 1205.6497}}].

\bibitem{Buttazzo:2013uya}
D.~Buttazzo, G.~Degrassi, P.~P. Giardino, G.~F. Giudice, F.~Sala, A.~Salvio,
  and A.~Strumia, {\it {Investigating the near-criticality of the Higgs
  boson}},  {\em JHEP} {\bf 12} (2013) 089,
  [\href{http://www.arxiv.org/abs/1307.3536}{{\tt 1307.3536}}].

\bibitem{Wilczek:1999be}
F.~Wilczek, {\it {Mass without mass. I: Most of matter}},  {\em Phys. Today}
  {\bf 52N11} (1999) 11--13.

\bibitem{Coleman:1973jx}
S.~R. Coleman and E.~J. Weinberg, {\it {Radiative Corrections as the Origin of
  Spontaneous Symmetry Breaking}},  {\em Phys. Rev.} {\bf D7} (1973)
  1888--1910.

\bibitem{Callan:1970yg}
C.~G. Callan, Jr., {\it {Broken scale invariance in scalar field theory}},
  {\em Phys. Rev.} {\bf D2} (1970) 1541--1547.

\bibitem{Symanzik:1970rt}
K.~Symanzik, {\it {Small distance behavior in field theory and power
  counting}},  {\em Commun. Math. Phys.} {\bf 18} (1970) 227--246.

\bibitem{Kronfeld:2012ym}
A.~S. Kronfeld, {\it {Lattice Gauge Theory and the Origin of Mass}},  in {\em
  100 Years of Subatomic Physics} (E.~M. Henley and S.~D. Ellis, eds.),
  pp.~493--518.
\newblock 2013.
\newblock \href{http://www.arxiv.org/abs/1209.3468}{{\tt 1209.3468}}.

\bibitem{Nambu:1960xd}
Y.~Nambu, {\it {Axial vector current conservation in weak interactions}},  {\em
  Phys. Rev. Lett.} {\bf 4} (1960) 380--382. [,107(1960)].

\bibitem{Nambu:1961tp}
Y.~Nambu and G.~Jona-Lasinio, {\it {Dynamical Model of Elementary Particles
  Based on an Analogy with Superconductivity. 1.}},  {\em Phys. Rev.} {\bf 122}
  (1961) 345--358. [,127(1961)].

\bibitem{Nambu:1961fr}
Y.~Nambu and G.~Jona-Lasinio, {\it {DYNAMICAL MODEL OF ELEMENTARY PARTICLES
  BASED ON AN ANALOGY WITH SUPERCONDUCTIVITY. II}},  {\em Phys. Rev.} {\bf 124}
  (1961) 246--254. [,141(1961)].

\bibitem{Hur:2011sv}
T.~Hur and P.~Ko, {\it {Scale invariant extension of the standard model with
  strongly interacting hidden sector}},  {\em Phys. Rev. Lett.} {\bf 106}
  (2011) 141802, [\href{http://www.arxiv.org/abs/1103.2571}{{\tt 1103.2571}}].

\bibitem{Heikinheimo:2013fta}
M.~Heikinheimo, A.~Racioppi, M.~Raidal, C.~Spethmann, and K.~Tuominen, {\it
  {Physical Naturalness and Dynamical Breaking of Classical Scale Invariance}},
   {\em Mod. Phys. Lett.} {\bf A29} (2014) 1450077,
  [\href{http://www.arxiv.org/abs/1304.7006}{{\tt 1304.7006}}].

\bibitem{Holthausen:2013ota}
M.~Holthausen, J.~Kubo, K.~S. Lim, and M.~Lindner, {\it {Electroweak and
  Conformal Symmetry Breaking by a Strongly Coupled Hidden Sector}},  {\em
  JHEP} {\bf 12} (2013) 076, [\href{http://www.arxiv.org/abs/1310.4423}{{\tt
  1310.4423}}].

\bibitem{Kubo:2014ova}
J.~Kubo, K.~S. Lim, and M.~Lindner, {\it {Electroweak Symmetry Breaking via
  QCD}},  {\em Phys. Rev. Lett.} {\bf 113} (2014) 091604,
  [\href{http://www.arxiv.org/abs/1403.4262}{{\tt 1403.4262}}].

\bibitem{Kubo:2015cna}
J.~Kubo and M.~Yamada, {\it {Genesis of electroweak and dark matter scales from
  a bilinear scalar condensate}},  {\em Phys. Rev.} {\bf D93} (2016), no.~7
  075016, [\href{http://www.arxiv.org/abs/1505.05971}{{\tt 1505.05971}}].

\bibitem{Hatanaka:2016rek}
H.~Hatanaka, D.-W. Jung, and P.~Ko, {\it {AdS/QCD approach to the
  scale-invariant extension of the standard model with a strongly interacting
  hidden sector}},  {\em JHEP} {\bf 08} (2016) 094,
  [\href{http://www.arxiv.org/abs/1606.02969}{{\tt 1606.02969}}].

\bibitem{Kubo:2014ida}
J.~Kubo, K.~S. Lim, and M.~Lindner, {\it {Gamma-ray Line from Nambu-Goldstone
  Dark Matter in a Scale Invariant Extension of the Standard Model}},  {\em
  JHEP} {\bf 09} (2014) 016, [\href{http://www.arxiv.org/abs/1405.1052}{{\tt
  1405.1052}}].

\bibitem{Ametani:2015jla}
Y.~Ametani, M.~Aoki, H.~Goto, and J.~Kubo, {\it {Nambu-Goldstone Dark Matter in
  a Scale Invariant Bright Hidden Sector}},  {\em Phys. Rev.} {\bf D91} (2015),
  no.~11 115007, [\href{http://www.arxiv.org/abs/1505.00128}{{\tt
  1505.00128}}].

\bibitem{Aghanim:2018eyx}
{\bf Planck} {\bf Collaboration}, N.~Aghanim {\em et~al.}, {\it {Planck 2018
  results. VI. Cosmological parameters}},
  \href{http://www.arxiv.org/abs/1807.06209}{{\tt 1807.06209}}.

\bibitem{Kubo:2018kho}
J.~Kubo, M.~Lindner, K.~Schmitz, and M.~Yamada, {\it {Planck mass and inflation
  as consequences of dynamically broken scale invariance}},  {\em Phys. Rev.}
  {\bf D100} (2019), no.~1 015037,
  [\href{http://www.arxiv.org/abs/1811.05950}{{\tt 1811.05950}}].

\bibitem{Bhattacharya:2014ara}
T.~Bhattacharya {\em et~al.}, {\it {QCD Phase Transition with Chiral Quarks and
  Physical Quark Masses}},  {\em Phys. Rev. Lett.} {\bf 113} (2014), no.~8
  082001, [\href{http://www.arxiv.org/abs/1402.5175}{{\tt 1402.5175}}].

\bibitem{Pisarski:1983ms}
R.~D. Pisarski and F.~Wilczek, {\it {Remarks on the Chiral Phase Transition in
  Chromodynamics}},  {\em Phys. Rev.} {\bf D29} (1984) 338--341.

\bibitem{DeTar:2009ef}
C.~DeTar and U.~M. Heller, {\it {QCD Thermodynamics from the Lattice}},  {\em
  Eur. Phys. J.} {\bf A41} (2009) 405--437,
  [\href{http://www.arxiv.org/abs/0905.2949}{{\tt 0905.2949}}]. [,1(2009)].

\bibitem{Meyer:2015wax}
H.~B. Meyer, {\it {QCD at non-zero temperature from the lattice}},  {\em PoS}
  {\bf LATTICE2015} (2016) 014,
  [\href{http://www.arxiv.org/abs/1512.06634}{{\tt 1512.06634}}].

\bibitem{Jin:2017jjp}
X.-Y. Jin, Y.~Kuramashi, Y.~Nakamura, S.~Takeda, and A.~Ukawa, {\it {Critical
  point phase transition for finite temperature 3-flavor QCD with
  non-perturbatively O($a$) improved Wilson fermions at $N_{\rm t}=10$}},  {\em
  Phys. Rev.} {\bf D96} (2017), no.~3 034523,
  [\href{http://www.arxiv.org/abs/1706.01178}{{\tt 1706.01178}}].

\bibitem{Kubo:2015joa}
J.~Kubo and M.~Yamada, {\it {Scale and electroweak first-order phase
  transitions}},  {\em PTEP} {\bf 2015} (2015), no.~9 093B01,
  [\href{http://www.arxiv.org/abs/1506.06460}{{\tt 1506.06460}}].

\bibitem{Aoki:2017aws}
M.~Aoki, H.~Goto, and J.~Kubo, {\it {Gravitational Waves from Hidden QCD Phase
  Transition}},  {\em Phys. Rev.} {\bf D96} (2017), no.~7 075045,
  [\href{http://www.arxiv.org/abs/1709.07572}{{\tt 1709.07572}}].

\bibitem{Schwaller:2015tja}
P.~Schwaller, {\it {Gravitational Waves from a Dark Phase Transition}},  {\em
  Phys. Rev. Lett.} {\bf 115} (2015), no.~18 181101,
  [\href{http://www.arxiv.org/abs/1504.07263}{{\tt 1504.07263}}].

\bibitem{Witten:1984rs}
E.~Witten, {\it {Cosmic Separation of Phases}},  {\em Phys. Rev.} {\bf D30}
  (1984) 272--285.

\bibitem{Maggiore:1999vm}
M.~Maggiore, {\it {Gravitational wave experiments and early universe
  cosmology}},  {\em Phys. Rept.} {\bf 331} (2000) 283--367,
  [\href{http://www.arxiv.org/abs/gr-qc/9909001}{{\tt gr-qc/9909001}}].

\bibitem{Binetruy:2012ze}
P.~Binetruy, A.~Bohe, C.~Caprini, and J.-F. Dufaux, {\it {Cosmological
  Backgrounds of Gravitational Waves and eLISA/NGO: Phase Transitions, Cosmic
  Strings and Other Sources}},  {\em JCAP} {\bf 1206} (2012) 027,
  [\href{http://www.arxiv.org/abs/1201.0983}{{\tt 1201.0983}}].

\bibitem{Seto:2003kc}
N.~Seto and J.~Yokoyama, {\it {Probing the equation of state of the early
  universe with a space laser interferometer}},  {\em J. Phys. Soc. Jap.} {\bf
  72} (2003) 3082--3086, [\href{http://www.arxiv.org/abs/gr-qc/0305096}{{\tt
  gr-qc/0305096}}].

\bibitem{Kuroyanagi:2008ye}
S.~Kuroyanagi, T.~Chiba, and N.~Sugiyama, {\it {Precision calculations of the
  gravitational wave background spectrum from inflation}},  {\em Phys. Rev.}
  {\bf D79} (2009) 103501, [\href{http://www.arxiv.org/abs/0804.3249}{{\tt
  0804.3249}}].

\bibitem{Schettler:2010dp}
S.~Schettler, T.~Boeckel, and J.~Schaffner-Bielich, {\it {Imprints of the QCD
  Phase Transition on the Spectrum of Gravitational Waves}},  {\em Phys. Rev.}
  {\bf D83} (2011) 064030, [\href{http://www.arxiv.org/abs/1010.4857}{{\tt
  1010.4857}}].

\bibitem{Schettler:2010wi}
T.~Boeckel, S.~Schettler, and J.~Schaffner-Bielich, {\it {The Cosmological QCD
  Phase Transition Revisited}},  {\em Prog. Part. Nucl. Phys.} {\bf 66} (2011)
  266--270, [\href{http://www.arxiv.org/abs/1012.3342}{{\tt 1012.3342}}].

\bibitem{Saikawa:2018rcs}
K.~Saikawa and S.~Shirai, {\it {Primordial gravitational waves, precisely: The
  role of thermodynamics in the Standard Model}},  {\em JCAP} {\bf 1805}
  (2018), no.~05 035, [\href{http://www.arxiv.org/abs/1803.01038}{{\tt
  1803.01038}}].

\bibitem{Hajkarim:2019csy}
F.~Hajkarim, J.~Schaffner-Bielich, S.~Wystub, and M.~M. Wygas, {\it {Effects of
  the QCD Equation of State and Lepton Asymmetry on Primordial Gravitational
  Waves}},  {\em Phys. Rev.} {\bf D99} (2019), no.~10 103527,
  [\href{http://www.arxiv.org/abs/1904.01046}{{\tt 1904.01046}}].

\bibitem{Abbott:2016blz}
{\bf LIGO Scientific, Virgo} {\bf Collaboration}, B.~P. Abbott {\em et~al.},
  {\it {Observation of Gravitational Waves from a Binary Black Hole Merger}},
  {\em Phys. Rev. Lett.} {\bf 116} (2016), no.~6 061102,
  [\href{http://www.arxiv.org/abs/1602.03837}{{\tt 1602.03837}}].

\bibitem{TheLIGOScientific:2017qsa}
{\bf LIGO Scientific, Virgo} {\bf Collaboration}, B.~P. Abbott {\em et~al.},
  {\it {GW170817: Observation of Gravitational Waves from a Binary Neutron Star
  Inspiral}},  {\em Phys. Rev. Lett.} {\bf 119} (2017), no.~16 161101,
  [\href{http://www.arxiv.org/abs/1710.05832}{{\tt 1710.05832}}].

\bibitem{GBM:2017lvd}
{\bf LIGO Scientific, Virgo, Fermi GBM, INTEGRAL, IceCube, AstroSat Cadmium
  Zinc Telluride Imager Team, IPN, Insight-Hxmt, ANTARES, Swift, AGILE Team,
  1M2H Team, Dark Energy Camera GW-EM, DES, DLT40, GRAWITA, Fermi-LAT, ATCA,
  ASKAP, Las Cumbres Observatory Group, OzGrav, DWF (Deeper Wider Faster
  Program), AST3, CAASTRO, VINROUGE, MASTER, J-GEM, GROWTH, JAGWAR,
  CaltechNRAO, TTU-NRAO, NuSTAR, Pan-STARRS, MAXI Team, TZAC Consortium, KU,
  Nordic Optical Telescope, ePESSTO, GROND, Texas Tech University, SALT Group,
  TOROS, BOOTES, MWA, CALET, IKI-GW Follow-up, H.E.S.S., LOFAR, LWA, HAWC,
  Pierre Auger, ALMA, Euro VLBI Team, Pi of Sky, Chandra Team at McGill
  University, DFN, ATLAS Telescopes, High Time Resolution Universe Survey,
  RIMAS, RATIR, SKA South Africa/MeerKAT} {\bf Collaboration}, B.~P. Abbott
  {\em et~al.}, {\it {Multi-messenger Observations of a Binary Neutron Star
  Merger}},  {\em Astrophys. J.} {\bf 848} (2017), no.~2 L12,
  [\href{http://www.arxiv.org/abs/1710.05833}{{\tt 1710.05833}}].

\bibitem{Konstandin:2011dr}
T.~Konstandin and G.~Servant, {\it {Cosmological Consequences of Nearly
  Conformal Dynamics at the TeV scale}},  {\em JCAP} {\bf 1112} (2011) 009,
  [\href{http://www.arxiv.org/abs/1104.4791}{{\tt 1104.4791}}].

\bibitem{Hashino:2015nxa}
K.~Hashino, S.~Kanemura, and Y.~Orikasa, {\it {Discriminative phenomenological
  features of scale invariant models for electroweak symmetry breaking}},  {\em
  Phys. Lett.} {\bf B752} (2016) 217--220,
  [\href{http://www.arxiv.org/abs/1508.03245}{{\tt 1508.03245}}].

\bibitem{Tsumura:2017knk}
K.~Tsumura, M.~Yamada, and Y.~Yamaguchi, {\it {Gravitational wave from dark
  sector with dark pion}},  {\em JCAP} {\bf 1707} (2017), no.~07 044,
  [\href{http://www.arxiv.org/abs/1704.00219}{{\tt 1704.00219}}].

\bibitem{Marzola:2017jzl}
L.~Marzola, A.~Racioppi, and V.~Vaskonen, {\it {Phase transition and
  gravitational wave phenomenology of scalar conformal extensions of the
  Standard Model}},  {\em Eur. Phys. J.} {\bf C77} (2017), no.~7 484,
  [\href{http://www.arxiv.org/abs/1704.01034}{{\tt 1704.01034}}].

\bibitem{Jinno:2016knw}
R.~Jinno and M.~Takimoto, {\it {Probing a classically conformal B-L model with
  gravitational waves}},  {\em Phys. Rev.} {\bf D95} (2017), no.~1 015020,
  [\href{http://www.arxiv.org/abs/1604.05035}{{\tt 1604.05035}}].

\bibitem{Prokopec:2018tnq}
T.~Prokopec, J.~Rezacek, and B.~Swiezewska, {\it {Gravitational waves from
  conformal symmetry breaking}},  {\em JCAP} {\bf 1902} (2019), no.~02 009,
  [\href{http://www.arxiv.org/abs/1809.11129}{{\tt 1809.11129}}].

\bibitem{Hashino:2018wee}
K.~Hashino, R.~Jinno, M.~Kakizaki, S.~Kanemura, T.~Takahashi, and M.~Takimoto,
  {\it {Selecting models of first-order phase transitions using the synergy
  between collider and gravitational-wave experiments}},  {\em Phys. Rev.} {\bf
  D99} (2019), no.~7 075011, [\href{http://www.arxiv.org/abs/1809.04994}{{\tt
  1809.04994}}].

\bibitem{Brdar:2018num}
V.~Brdar, A.~J. Helmboldt, and J.~Kubo, {\it {Gravitational Waves from
  First-Order Phase Transitions: LIGO as a Window to Unexplored Seesaw
  Scales}},  {\em JCAP} {\bf 1902} (2019) 021,
  [\href{http://www.arxiv.org/abs/1810.12306}{{\tt 1810.12306}}].

\bibitem{Miura:2018dsy}
K.~Miura, H.~Ohki, S.~Otani, and K.~Yamawaki, {\it {Gravitational Waves from
  Walking Technicolor}},  \href{http://www.arxiv.org/abs/1811.05670}{{\tt
  1811.05670}}.

\bibitem{Marzo:2018nov}
C.~Marzo, L.~Marzola, and V.~Vaskonen, {\it {Phase transition and vacuum
  stability in the classically conformal B-L model}},  {\em Eur. Phys. J.} {\bf
  C79} (2019), no.~7 601, [\href{http://www.arxiv.org/abs/1811.11169}{{\tt
  1811.11169}}].

\bibitem{Croon:2019iuh}
D.~Croon, R.~Houtz, and V.~Sanz, {\it {Dynamical Axions and Gravitational
  Waves}},  {\em JHEP} {\bf 07} (2019) 146,
  [\href{http://www.arxiv.org/abs/1904.10967}{{\tt 1904.10967}}].

\bibitem{Dai:2019ksi}
D.-C. Dai and D.~Stojkovic, {\it {Primordial scalar gravitational waves
  produced at the QCD phase transition due to the trace anomaly}},  {\em Class.
  Quant. Grav.} {\bf 36} (2019), no.~14 145004,
  [\href{http://www.arxiv.org/abs/1905.05850}{{\tt 1905.05850}}].

\bibitem{Mohamadnejad:2019vzg}
A.~Mohamadnejad, {\it {Gravitational waves from scale-invariant vector dark
  matter model: Probing below the neutrino-floor}},
  \href{http://www.arxiv.org/abs/1907.08899}{{\tt 1907.08899}}.

\bibitem{Kosowsky:1991ua}
A.~Kosowsky, M.~S. Turner, and R.~Watkins, {\it {Gravitational radiation from
  colliding vacuum bubbles}},  {\em Phys. Rev.} {\bf D45} (1992) 4514--4535.

\bibitem{Kosowsky:1992rz}
A.~Kosowsky, M.~S. Turner, and R.~Watkins, {\it {Gravitational waves from first
  order cosmological phase transitions}},  {\em Phys. Rev. Lett.} {\bf 69}
  (1992) 2026--2029.

\bibitem{Kosowsky:1992vn}
A.~Kosowsky and M.~S. Turner, {\it {Gravitational radiation from colliding
  vacuum bubbles: envelope approximation to many bubble collisions}},  {\em
  Phys. Rev.} {\bf D47} (1993) 4372--4391,
  [\href{http://www.arxiv.org/abs/astro-ph/9211004}{{\tt astro-ph/9211004}}].

\bibitem{Helmboldt:2019pan}
A.~J. Helmboldt, J.~Kubo, and S.~van~der Woude, {\it {Observational prospects
  for gravitational waves from hidden or dark chiral phase transitions}},
  \href{http://www.arxiv.org/abs/1904.07891}{{\tt 1904.07891}}.

\bibitem{Hogan:1984hx}
C.~J. Hogan, {\it {NUCLEATION OF COSMOLOGICAL PHASE TRANSITIONS}},  {\em Phys.
  Lett.} {\bf 133B} (1983) 172--176.

\bibitem{Falkowski:2015iwa}
A.~Falkowski, C.~Gross, and O.~Lebedev, {\it {A second Higgs from the Higgs
  portal}},  {\em JHEP} {\bf 05} (2015) 057,
  [\href{http://www.arxiv.org/abs/1502.01361}{{\tt 1502.01361}}].

\bibitem{Robens:2016xkb}
T.~Robens and T.~Stefaniak, {\it {LHC Benchmark Scenarios for the Real Higgs
  Singlet Extension of the Standard Model}},  {\em Eur. Phys. J.} {\bf C76}
  (2016), no.~5 268, [\href{http://www.arxiv.org/abs/1601.07880}{{\tt
  1601.07880}}].

\bibitem{Linde:1981zj}
A.~D. Linde, {\it {Decay of the False Vacuum at Finite Temperature}},  {\em
  Nucl. Phys.} {\bf B216} (1983) 421. [Erratum: Nucl. Phys.B223,544(1983)].

\bibitem{Wainwright:2011kj}
C.~L. Wainwright, {\it {CosmoTransitions: Computing Cosmological Phase
  Transition Temperatures and Bubble Profiles with Multiple Fields}},  {\em
  Comput. Phys. Commun.} {\bf 183} (2012) 2006--2013,
  [\href{http://www.arxiv.org/abs/1109.4189}{{\tt 1109.4189}}].

\bibitem{Hindmarsh:2017gnf}
M.~Hindmarsh, S.~J. Huber, K.~Rummukainen, and D.~J. Weir, {\it {Shape of the
  acoustic gravitational wave power spectrum from a first order phase
  transition}},  {\em Phys. Rev.} {\bf D96} (2017), no.~10 103520,
  [\href{http://www.arxiv.org/abs/1704.05871}{{\tt 1704.05871}}].

\bibitem{Ellis:2018mja}
J.~Ellis, M.~Lewicki, and J.~M. No, {\it {On the Maximal Strength of a
  First-Order Electroweak Phase Transition and its Gravitational Wave Signal}},
   \href{http://www.arxiv.org/abs/1809.08242}{{\tt 1809.08242}}.
  [JCAP1904,003(2019)].

\bibitem{Ellis:2019oqb}
J.~Ellis, M.~Lewicki, J.~M. No, and V.~Vaskonen, {\it {Gravitational wave
  energy budget in strongly supercooled phase transitions}},  {\em JCAP} {\bf
  1906} (2019), no.~06 024, [\href{http://www.arxiv.org/abs/1903.09642}{{\tt
  1903.09642}}].

\bibitem{Espinosa:2010hh}
J.~R. Espinosa, T.~Konstandin, J.~M. No, and G.~Servant, {\it {Energy Budget of
  Cosmological First-order Phase Transitions}},  {\em JCAP} {\bf 1006} (2010)
  028, [\href{http://www.arxiv.org/abs/1004.4187}{{\tt 1004.4187}}].

\bibitem{Thrane:2013oya}
E.~Thrane and J.~D. Romano, {\it {Sensitivity curves for searches for
  gravitational-wave backgrounds}},  {\em Phys. Rev.} {\bf D88} (2013), no.~12
  124032, [\href{http://www.arxiv.org/abs/1310.5300}{{\tt 1310.5300}}].

\bibitem{Seto:2001qf}
N.~Seto, S.~Kawamura, and T.~Nakamura, {\it {Possibility of direct measurement
  of the acceleration of the universe using 0.1-Hz band laser interferometer
  gravitational wave antenna in space}},  {\em Phys. Rev. Lett.} {\bf 87}
  (2001) 221103, [\href{http://www.arxiv.org/abs/astro-ph/0108011}{{\tt
  astro-ph/0108011}}].

\bibitem{Kawamura:2006up}
S.~Kawamura {\em et~al.}, {\it {The Japanese space gravitational wave antenna
  DECIGO}},  {\em Class. Quant. Grav.} {\bf 23} (2006) S125--S132.

\bibitem{Kawamura:2011zz}
S.~Kawamura {\em et~al.}, {\it {The Japanese space gravitational wave antenna:
  DECIGO}},  {\em Class. Quant. Grav.} {\bf 28} (2011) 094011.

\bibitem{BBO}
S.~Phinney {\em et~al.}, {\it {The Big Bang Observer: Direct detection of
  gravitational waves from the birth of the Universe to the Present}},  {\em
  NASA Mission Concept Study (2004)}.

\bibitem{Cornish:2005qw}
N.~J. Cornish and J.~Crowder, {\it {LISA data analysis using MCMC methods}},
  {\em Phys. Rev.} {\bf D72} (2005) 043005,
  [\href{http://www.arxiv.org/abs/gr-qc/0506059}{{\tt gr-qc/0506059}}].

\bibitem{Corbin:2005ny}
V.~Corbin and N.~J. Cornish, {\it {Detecting the cosmic gravitational wave
  background with the big bang observer}},  {\em Class. Quant. Grav.} {\bf 23}
  (2006) 2435--2446, [\href{http://www.arxiv.org/abs/gr-qc/0512039}{{\tt
  gr-qc/0512039}}].

\bibitem{Kunihiro:1983ej}
T.~Kunihiro and T.~Hatsuda, {\it {A Selfconsistent Mean Field Approach to the
  Dynamical Symmetry Breaking: The Effective Potential of the
  {Nambu-Jona-Lasinio} Model}},  {\em Prog. Theor. Phys.} {\bf 71} (1984) 1332.

\bibitem{Hatsuda:1994pi}
T.~Hatsuda and T.~Kunihiro, {\it {QCD phenomenology based on a chiral effective
  Lagrangian}},  {\em Phys. Rept.} {\bf 247} (1994) 221--367,
  [\href{http://www.arxiv.org/abs/hep-ph/9401310}{{\tt hep-ph/9401310}}].

\bibitem{Apreda:2001us}
R.~Apreda, M.~Maggiore, A.~Nicolis, and A.~Riotto, {\it {Gravitational waves
  from electroweak phase transitions}},  {\em Nucl. Phys.} {\bf B631} (2002)
  342--368, [\href{http://www.arxiv.org/abs/gr-qc/0107033}{{\tt
  gr-qc/0107033}}].

\bibitem{Kamionkowski:1993fg}
M.~Kamionkowski, A.~Kosowsky, and M.~S. Turner, {\it {Gravitational radiation
  from first order phase transitions}},  {\em Phys. Rev.} {\bf D49} (1994)
  2837--2851, [\href{http://www.arxiv.org/abs/astro-ph/9310044}{{\tt
  astro-ph/9310044}}].

\bibitem{Caprini:2007xq}
C.~Caprini, R.~Durrer, and G.~Servant, {\it {Gravitational wave generation from
  bubble collisions in first-order phase transitions: An analytic approach}},
  {\em Phys. Rev.} {\bf D77} (2008) 124015,
  [\href{http://www.arxiv.org/abs/0711.2593}{{\tt 0711.2593}}].

\bibitem{Huber:2008hg}
S.~J. Huber and T.~Konstandin, {\it {Gravitational Wave Production by
  Collisions: More Bubbles}},  {\em JCAP} {\bf 0809} (2008) 022,
  [\href{http://www.arxiv.org/abs/0806.1828}{{\tt 0806.1828}}].

\bibitem{Hindmarsh:2013xza}
M.~Hindmarsh, S.~J. Huber, K.~Rummukainen, and D.~J. Weir, {\it {Gravitational
  waves from the sound of a first order phase transition}},  {\em Phys. Rev.
  Lett.} {\bf 112} (2014) 041301,
  [\href{http://www.arxiv.org/abs/1304.2433}{{\tt 1304.2433}}].

\bibitem{Hindmarsh:2015qta}
M.~Hindmarsh, S.~J. Huber, K.~Rummukainen, and D.~J. Weir, {\it {Numerical
  simulations of acoustically generated gravitational waves at a first order
  phase transition}},  {\em Phys. Rev.} {\bf D92} (2015), no.~12 123009,
  [\href{http://www.arxiv.org/abs/1504.03291}{{\tt 1504.03291}}].

\bibitem{Giblin:2013kea}
J.~T. Giblin, Jr. and J.~B. Mertens, {\it {Vacuum Bubbles in the Presence of a
  Relativistic Fluid}},  {\em JHEP} {\bf 12} (2013) 042,
  [\href{http://www.arxiv.org/abs/1310.2948}{{\tt 1310.2948}}].

\bibitem{Giblin:2014qia}
J.~T. Giblin and J.~B. Mertens, {\it {Gravitional radiation from first-order
  phase transitions in the presence of a fluid}},  {\em Phys. Rev.} {\bf D90}
  (2014), no.~2 023532, [\href{http://www.arxiv.org/abs/1405.4005}{{\tt
  1405.4005}}].

\bibitem{Kosowsky:2001xp}
A.~Kosowsky, A.~Mack, and T.~Kahniashvili, {\it {Gravitational radiation from
  cosmological turbulence}},  {\em Phys. Rev.} {\bf D66} (2002) 024030,
  [\href{http://www.arxiv.org/abs/astro-ph/0111483}{{\tt astro-ph/0111483}}].

\bibitem{Caprini:2006jb}
C.~Caprini and R.~Durrer, {\it {Gravitational waves from stochastic
  relativistic sources: Primordial turbulence and magnetic fields}},  {\em
  Phys. Rev.} {\bf D74} (2006) 063521,
  [\href{http://www.arxiv.org/abs/astro-ph/0603476}{{\tt astro-ph/0603476}}].

\bibitem{Kahniashvili:2008pe}
T.~Kahniashvili, L.~Campanelli, G.~Gogoberidze, Y.~Maravin, and B.~Ratra, {\it
  {Gravitational Radiation from Primordial Helical Inverse Cascade MHD
  Turbulence}},  {\em Phys. Rev.} {\bf D78} (2008) 123006,
  [\href{http://www.arxiv.org/abs/0809.1899}{{\tt 0809.1899}}]. [Erratum: Phys.
  Rev.D79,109901(2009)].

\bibitem{Kahniashvili:2008pf}
T.~Kahniashvili, A.~Kosowsky, G.~Gogoberidze, and Y.~Maravin, {\it
  {Detectability of Gravitational Waves from Phase Transitions}},  {\em Phys.
  Rev.} {\bf D78} (2008) 043003,
  [\href{http://www.arxiv.org/abs/0806.0293}{{\tt 0806.0293}}].

\bibitem{Kahniashvili:2009mf}
T.~Kahniashvili, L.~Kisslinger, and T.~Stevens, {\it {Gravitational Radiation
  Generated by Magnetic Fields in Cosmological Phase Transitions}},  {\em Phys.
  Rev.} {\bf D81} (2010) 023004,
  [\href{http://www.arxiv.org/abs/0905.0643}{{\tt 0905.0643}}].

\bibitem{Caprini:2009yp}
C.~Caprini, R.~Durrer, and G.~Servant, {\it {The stochastic gravitational wave
  background from turbulence and magnetic fields generated by a first-order
  phase transition}},  {\em JCAP} {\bf 0912} (2009) 024,
  [\href{http://www.arxiv.org/abs/0909.0622}{{\tt 0909.0622}}].

\bibitem{Kisslinger:2015hua}
L.~Kisslinger and T.~Kahniashvili, {\it {Polarized Gravitational Waves from
  Cosmological Phase Transitions}},  {\em Phys. Rev.} {\bf D92} (2015), no.~4
  043006, [\href{http://www.arxiv.org/abs/1505.03680}{{\tt 1505.03680}}].

\bibitem{Coleman:1977py}
S.~R. Coleman, {\it {The Fate of the False Vacuum. 1. Semiclassical Theory}},
  {\em Phys. Rev.} {\bf D15} (1977) 2929--2936. [Erratum: Phys.
  Rev.D16,1248(1977)].

\bibitem{Callan:1977pt}
C.~G. Callan, Jr. and S.~R. Coleman, {\it {The Fate of the False Vacuum. 2.
  First Quantum Corrections}},  {\em Phys. Rev.} {\bf D16} (1977) 1762--1768.

\bibitem{Jinno:2019bxw}
R.~Jinno, T.~Konstandin, and M.~Takimoto, {\it {Relativistic bubble collisions
  -- a closer look}},  \href{http://www.arxiv.org/abs/1906.02588}{{\tt
  1906.02588}}.

\bibitem{Bodeker:2009qy}
D.~Bodeker and G.~D. Moore, {\it {Can electroweak bubble walls run away?}},
  {\em JCAP} {\bf 0905} (2009) 009,
  [\href{http://www.arxiv.org/abs/0903.4099}{{\tt 0903.4099}}].

\bibitem{Moore:2014lga}
C.~J. Moore, R.~H. Cole, and C.~P.~L. Berry, {\it {Gravitational-wave
  sensitivity curves}},  {\em Class. Quant. Grav.} {\bf 32} (2015), no.~1
  015014, [\href{http://www.arxiv.org/abs/1408.0740}{{\tt 1408.0740}}].

\bibitem{Yagi:2011yu}
K.~Yagi, N.~Tanahashi, and T.~Tanaka, {\it {Probing the size of extra dimension
  with gravitational wave astronomy}},  {\em Phys. Rev.} {\bf D83} (2011)
  084036, [\href{http://www.arxiv.org/abs/1101.4997}{{\tt 1101.4997}}].

\bibitem{Yagi:2013du}
K.~Yagi, {\it {Scientific Potential of DECIGO Pathfinder and Testing GR with
  Space-Borne Gravitational Wave Interferometers}},  {\em Int. J. Mod. Phys.}
  {\bf D22} (2013) 1341013, [\href{http://www.arxiv.org/abs/1302.2388}{{\tt
  1302.2388}}].

\bibitem{Isoyama:2018rjb}
S.~Isoyama, H.~Nakano, and T.~Nakamura, {\it {Multiband Gravitational-Wave
  Astronomy: Observing binary inspirals with a decihertz detector, B-DECIGO}},
  {\em PTEP} {\bf 2018} (2018), no.~7 073E01,
  [\href{http://www.arxiv.org/abs/1802.06977}{{\tt 1802.06977}}].

\bibitem{Alanne:2019bsm}
T.~Alanne, T.~Hugle, M.~Platscher, and K.~Schmitz, {\it {A fresh look at the
  gravitational-wave signal from cosmological phase transitions}},
  \href{http://www.arxiv.org/abs/1909.11356}{{\tt 1909.11356}}.

\bibitem{Brower:2019oor}
{\bf USQCD} {\bf Collaboration}, R.~C. Brower, A.~Hasenfratz, E.~T. Neil,
  S.~Catterall, G.~Fleming, J.~Giedt, E.~Rinaldi, D.~Schaich, E.~Weinberg, and
  O.~Witzel, {\it {Lattice Gauge Theory for Physics Beyond the Standard
  Model}},  \href{http://www.arxiv.org/abs/1904.09964}{{\tt 1904.09964}}.

\end{thebibliography}\endgroup

\end{document}